\documentclass[cqg,floatfix,superscriptaddress,twocolumn,nofootinbib]{revtex4}

\usepackage{amssymb,amsmath}
\usepackage{graphicx}
\usepackage[percent]{overpic}
\usepackage{overpic}
\usepackage{xcolor}

\begin{document}  

\title{Tidal deformability of boson stars and dark matter clumps} 

\author{Raissa F. P. Mendes}
\affiliation{Instituto de F\'isica, Universidade Federal Fluminense, \\
Av.~Gal.~Milton Tavares de Souza s/n, Gragoat\'a, 24210-346 Niter\'oi, Rio de Janeiro, Brazil.}

\author{Huan Yang}
\affiliation{Department of Physics, Princeton University, Princeton, New Jersey 08544, USA.}

\date{\today}

\begin{abstract} 
In this work we consider minimally-coupled boson stars immersed in a tidal environment and compute their tidal deformability to leading order. We also describe an approximate correspondence between Newtonian boson star configurations (described by the Schr\"{o}dinger-Poisson equations) and dynamical dark matter clumps (described by the collisionless Boltzmann equation). This allows us to map our results for the tidal deformability of boson stars to approximate statements for dark matter clumps. \end{abstract}

\maketitle 

\section{Introduction}

\subsection{Deformed boson stars}

Boson stars are self-gravitating star-like objects consisting of one or multiple scalar fields \cite{Jetzer1992,Schunck2003,Liebling2012}. Depending on the nature of the scalar self-interaction and its coupling to gravity, these objects have been proposed as ultracompact candidates that could mimic the gravitational-wave emission from black holes in compact binary mergers \cite{PhysRevLett.116.171101}; as dark matter candidates \cite{Sin1994,Lee1996}; and often as convenient toy models for matter in strongly curved spacetimes. Examples of the last scenario include the pioneer work of Detweiler in studying black hole super-radiance \cite{PhysRevD.22.2323}, ultra-relativistic boson star collisions to test the hoop conjecture by Choptuik and Pretorius \cite{PhysRevLett.104.111101}, among others \cite{Liebling2012}.

The primary goal of the present work is to study the tidal deformability of boson stars immersed in an external tidal environment and to compute their leading order response to the applied tidal field. This is encoded in the so-called tidal deformabilities (or tidal Love numbers). Such understanding is useful as it is directly relevant to the dynamics and gravitational wave emission of compact boson stars in binaries. In particular, here we compute the $l=2$ tidal Love number of minimally-coupled boson stars. Our calculation is fully relativistic, since highly compact boson stars have many phenomenological applications (e.g. as black hole mimickers), but we also describe in detail the Newtonian limit that is relevant for most dark matter models and for the discussion in the second half of the paper. A subsequent study by Cardoso \textit{et al.}  \cite{cardoso2017testing} greatly extended this investigation to self-interacting scalar fields and to higher-multipole Love numbers (see also \cite{sennett2017} for a subsequent related work).

Gearing up with the understanding of deformed boson stars, we then explore an approximate mapping between boson star solutions and possible compact configurations of dark matter, as explained briefly below.

\subsection{Connection to dark matter clumps}

Dark matter is one of the main pillars of the current standard cosmological model, comprising approximately $83\%$ of all matter within the universe and playing a crucial role in structure and galaxy formation. Although the first evidences for dark matter were found as early as the 1930s \cite{Zwicky1933}, its nature is still elusive, with theoretical candidates ranging from weakly interacting massive particles with masses in the GeV range \cite{PhysRevD.86.023506} to ultralight scalar particles with masses as low as $10^{-22}$eV \cite{Hu2000}.
Given the uncertainty about the fundamental properties of dark matter, experimental and observational efforts must be multifaceted, and many are already probing mass and cross-section regimes relevant for theoretical models (see, e.g., the recent reports \cite{LUX,PandaXII,Xenon100,PhysRevLett.115.231301}).

If dark matter can cluster into compact configurations, an interesting possibility would be to probe its properties with gravitational wave measurements. This would complement current searches for byproducts of dark matter annihilation or decay (e.g., galaxy gamma-ray excess, neutrinos and cosmic rays \cite{Gaskins:2016}), by probing the only interaction guaranteed to affect dark matter, i.e.~gravity. 
With the recent direct detections of gravitational waves by Advanced LIGO \cite{PhysRevLett.116.061102,PhysRevLett.116.241103}, and successful launch of LISA Pathfinder \cite{PhysRevLett.116.231101}, it seems timely to investigate the possibility of using gravitational wave measurements to study dark matter physics.

Along this line, a second aim of this work is to describe an approximate mapping between Newtonian boson star configurations and particle dark matter clumps. Boson stars, in the Newtonian limit, are described by a wave function satisfying the Schr\"{o}dinger-Poisson equations but also admit a phase space representation, e.g., via the Wigner function. The mapping we discuss relies on the similarities between the collisionless Boltzmann equation for the phase-space density distribution describing dark matter particles, and the equation for the Wigner function in a certain limit. The basic idea dates back to the work of Widrow and Kaiser, who proposed an approximate evolution scheme for the phase-space density distribution relying on a similar wave-mechanical approach \cite{widrow1993using}. 
This procedure was extended by Schaller {\it et al.}~\cite{schaller2014new} and by Widrow \cite{Widrow1997}, and was validated against $N$-body simulations in \cite{widrow1993using,Davies1997}, where it was shown to have a comparable or reduced computational cost with respect to the more standard approaches. Here, we adapt the construction of Refs.~\cite{widrow1993using,schaller2014new,Davies1997} to our perturbative setup and discuss how to map our results for the tidal deformability of boson stars to similar (approximate) statements about the dynamical response of a collisionless dark matter clump under an external gravitational field. 

\subsection{Organization of the paper}

The organization of this paper is as follows. In Sec.~\ref{sec2} we present our main results on the tidal deformability of a minimally coupled boson star. In Sec.~\ref{sec3} we introduce and discuss an approximate mapping between solutions of the Schr\"{o}dinger-Poisson equations and solutions of the collisionless Boltzmann equation, emphasizing the assumptions involved and the domain of validity of the approximation. In Sec.~\ref{sec4}, we discuss how this correspondence allows us to estimate the tidal deformability of a collisionless dark matter clump from our results for boson stars. We conclude in Sec.~\ref{sec5}. Additional support for our claims is presented in the Appendices. Throughout most of the paper, we adopt natural units such that $c=G=1$, but recover $c$'s and $G$'s at certain points for the sake of clarity.

%%%%%%%%%%%%%%%%%%%%%%%%%%%%%%%%%%%%%%%%%%%%%
\section{Boson stars in a tidal environment}\label{sec2}

In this section, we analyze the leading order response of a boson star to an applied tidal field. Here we present the fully relativistic calculation, discussing later the appropriate Newtonian limit, which is relevant in dark matter models. 

We consider a massive (complex) scalar field minimally coupled to gravity and with no self-interactions, described by the action
\begin{equation}\label{eq:action_bs}
S = \int{d^4 x \sqrt{-g} \left[ \frac{R}{16 \pi} - \nabla_\alpha \Phi^* \nabla^\alpha \Phi - \mu^2 |\Phi|^2 \right]},
\end{equation}
where $\mu := m/\hbar$ is a mass parameter with units of inverse length.
The field equations that follow from Eq.~(\ref{eq:action_bs}) are the Einstein,
\begin{align}\label{eq:EE}
G_{\alpha\beta} = 8\pi & \left[ \nabla_\alpha \Phi^* \nabla_\beta \Phi + \nabla_\beta \Phi^* \nabla_\alpha \Phi \right. \nonumber \\
{}& \left. - g_{\alpha\beta} \left(\nabla_\rho \Phi^* \nabla^\rho \Phi + \mu^2 |\Phi|^2 \right) \right],
\end{align}
and Klein-Gordon,
\begin{equation}\label{eq:KG}
\nabla^\alpha \nabla_\alpha \Phi = \mu^2 \Phi,
\end{equation}
equations, together with the complex conjugate of Eq.~(\ref{eq:KG}). 

In the following, we first construct spherically symmetric equilibrium solutions of Eqs.~(\ref{eq:EE}) and (\ref{eq:KG}) and then consider perturbations to these solutions (in the Regge-Wheeler gauge), sourced by an external tidal field. 
We assume the tidal field to be nearly static or, equivalently, that the boson star response to a change in the tidal environment is adiabatic. This assumption is appropriate for tidal interactions that occur over an external time scale that is long compared with the one associated with the internal dynamics of the boson star. 

The external, nearly static tidal field is characterized by the quadrupole tidal moment $\mathcal{E}_{ij}$. In a Newtonian setting, $\mathcal{E}_{ij} := \partial_i \partial_j U_\textrm{ext}$, where $U_\textrm{ext}$ is the external potential (which is evaluated at the body's center of mass after differentiation) \cite{poisson:gravity}. In response to the applied tidal field, the star develops a nonzero quadrupole moment $Q_{ij}$ which, to linear order, is proportional to $\mathcal{E}_{ij}$, $Q_{ij} = -\lambda \mathcal{E}_{ij}$. More specifically, in the star's local rest frame, the metric component $g_{tt}$ can be written to leading order as \cite{Thorne1998}
\begin{equation} \label{eq:asymp_gtt}
g_{tt} = -1 + \frac{2M}{r} - \mathcal{E}_{ij} x^i x^j \left( 1 + \frac{3 \lambda}{r^5} \right),
\end{equation} 
where $M$ is the star's gravitational mass and $x^i$ defines a Cartesian coordinate system, with $r^2:=\delta_{jk} x^j x^k$. Equation (\ref{eq:asymp_gtt}) includes the external quadrupolar tidal potential and the leading order response from the body. 
The coefficient $\lambda$ gives a measure of the deformability of the boson star, and this is the quantity we ultimately want to compute.

\subsection{Background}
\label{sec:background}

We choose the unperturbed configuration to be spherically symmetric and static, in which case the spacetime line element can be written as
\begin{equation} \label{eq:bmetric}
ds^2 = -e^{v(r)} dt^2 + e^{u(r)} dr^2 + r^2 d \theta^2 + r^2 \sin^2 \theta d \varphi^2 \,.
\end{equation}
The scalar field is taken to have a harmonic time dependence, 
\begin{equation}\label{eq:bfield}
\Phi_0 (t,r) = e^{-i\omega t} \phi_0 (r),
\end{equation} 
which is consistent with a static gravitational field. With these ansatzes, the field equations (\ref{eq:EE}) and (\ref{eq:KG}) reduce to the following set of ordinary differential equations:
\begin{subequations}\label{eq:back}
\begin{align}
u' & = \frac{1 - e^u}{r} + 8\pi r e^u [(\omega^2 e^{-v} \!+\! \mu^2) \phi^2_0 + e^{-u} \phi_0^{\prime 2}]\,, \\
v' & = \frac{e^u - 1}{r} +8 \pi r e^u [(\omega^2 e^{-v} \!-\! \mu^2) \phi^2_0 + e^{-u} \phi_0^{\prime 2}]\,, \\
\phi''_0 & + \left ( \frac{2}{r} + \frac{v'-u'}{2} \right )\phi'_0 + e^u (\omega^2 e^{-v} - \mu^2)\phi_0 = 0 \,, 
\end{align}
\end{subequations}
with a prime denoting a radial derivative. Note that it is possible to define dimensionless variables $\tilde{r} = \mu r$ and $\tilde{\omega} = \omega/\mu$, in terms of which the equations above are not explicitly $\mu$-dependent.

Equations (\ref{eq:back}) must be solved subject to regularity conditions at $r=0$, namely $u(0) = 0$ and $\phi'(0) = 0$, and boundary conditions at spatial infinity:
\begin{equation}
\lim_{r\to \infty} v(r) = 0, \qquad \lim_{r\to \infty} \phi_0(r) = 0.
\end{equation}
For a given value of $\phi_c := \phi_0(0)$, solutions satisfying these boundary conditions exist only for a discrete set of eigenfrequencies $\omega$. They are computed numerically by a shooting algorithm, which we briefly describe. 
We evolve the system (\ref{eq:back}) from $r=0$, with initial conditions $u(0)=0$, $v(0)=v_c$, $\phi_0(0) = \phi_c$, and $\phi_0'(0) = 0$, where $v_c$ is in principle arbitrary\footnote{In practice, the numerical integration starts at some small value $r = \epsilon$, where appropriate boundary conditions are obtained from a Taylor expansion of the solution around $r=0$.}. For a given value of $\phi_c>0$, the solution thus obtained for $\phi_0(r)$ diverges to $+\infty$ when $\omega \approx 0$; then, as we increase the value of $\omega$, to $- \infty$ after crossing zero once, then to $+ \infty$ after crossing zero twice, and so forth. Solutions with vanishing scalar field at spatial infinity, and different number of nodes, exist for frequencies $\omega$ in the threshold between these various asymptotic behaviours. We are interested in the ground-state solution, which has no nodes. By fine-tuning the value of $\omega$, we can obtain an approximation to this solution, which crosses zero at increasingly higher values of $r$. The numerical solution thus obtained is in general such that $v_\infty := \lim_{r\to\infty} v(r) \neq 0$. However, since Eqs.~(\ref{eq:back}) are invariant under a constant shift in $v(r)$, $v \to v + v_0$, accompanied by a rescaling in frequency, $\omega \to e^{v_0/2} \omega$, the actual solution, satisfying the correct boundary conditions, can be obtained a posteriori by $v(r) \to v(r) - v_\infty$, $\omega \to e^{-v_\infty/2} \omega$.

\begin{figure}[thb]
\includegraphics[width=8.4cm]{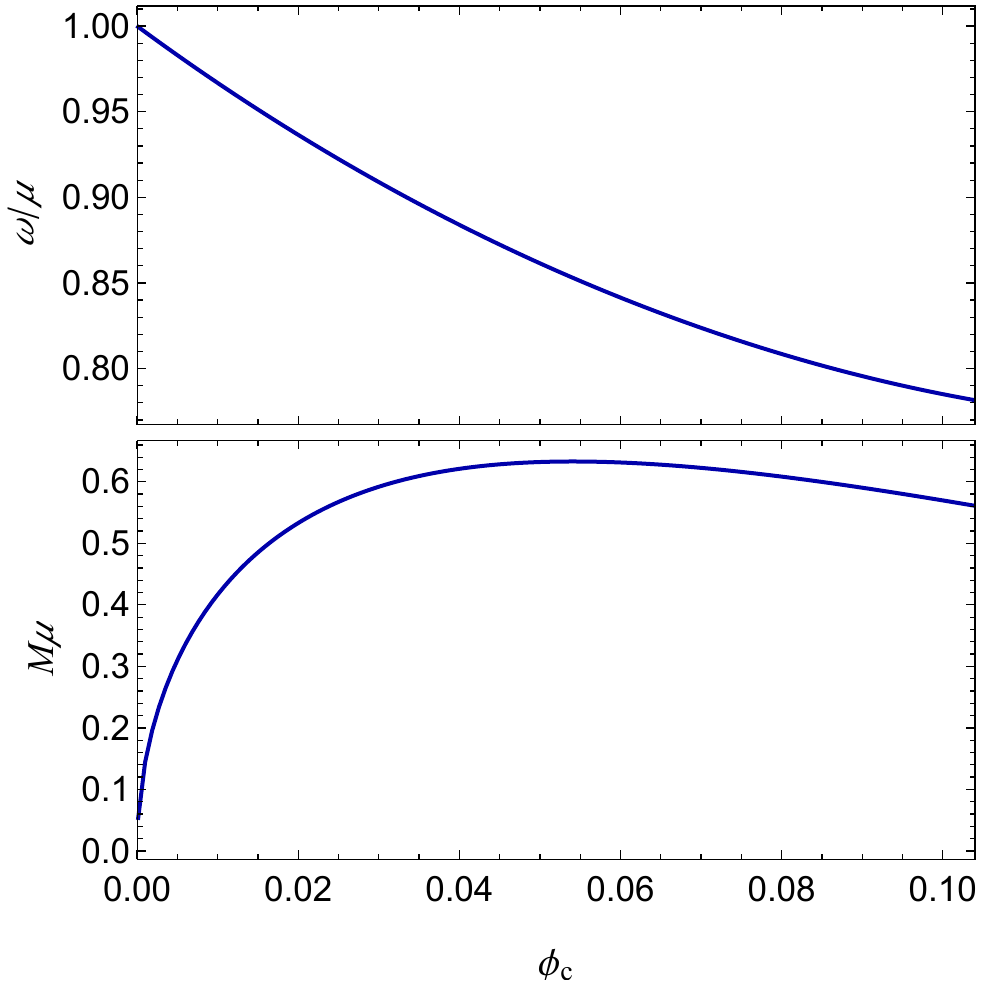}
\caption{Sequence of equilibrium boson star configurations, labelled by value $\phi_c$ of the central scalar field. The frequency $\omega$ and the total mass $M$ are shown as a function of $\phi_c$.}
\label{fig:background}
\end{figure}

By varying the central value of the scalar field, we obtain a sequence of boson star configurations, as shown in Fig.~\ref{fig:background}. In the upper panel, the frequency $\omega$, obtained through the shooting procedure described above, is shown as a function of $\phi_c$. We notice that $\omega \to \mu$ as $\phi_c \to 0$, as characteristic of the Newtonian limit. 
The gravitational mass of the star is defined as $M = \lim_{r\to\infty} m(r)$, where $m(r):= (r/2) (1- e^{-u(r)})$ and is plotted in the lower panel of Fig.~\ref{fig:background}. As for fluid stars, we see a turning point in the $M$-$\phi_c$ diagram, signalling a change in stability of the equilibrium configurations \cite{Jetzer1992}. Stable configurations exist for $\phi_c \lesssim 0.054$.

Finally, note that, since the scalar field profile decays exponentially for $r\gg 1/\mu$, it is possible to define an effective notion of radius of the boson star \cite{Schunck2003}; for example, we can define the effective radius $R$ as being such that $m(R) = 0.99 M$.

\subsection{Perturbations}
\label{sec:perturbations}

In order to determine the response of a boson star to an applied tidal field, we first consider general perturbations of the equilibrium configuration. We write
\begin{equation}
g_{\alpha\beta} = g^{(0)}_{\alpha\beta} + h_{\alpha\beta},
\qquad 
\Phi = \Phi_0 + \delta \Phi, 
\end{equation}
where $g^{(0)}_{\alpha\beta}$ is given by Eq.~(\ref{eq:bmetric}) and $\Phi_0$ is given by Eq.~(\ref{eq:bfield}), and work in first order in the perturbed quantities. We adopt the Regge-Wheeler gauge and restrict consideration to quadrupolar ($l=2$) even-parity static perturbations of the metric, namely,
\begin{align}\label{eq:metric_perturbation}
&h_{\alpha \beta} = \sum_{{\rm m}=-2}^{2} Y_{2 {\rm m}}(\theta,\varphi) \times \nonumber \\
&{\rm diag} [-e^{v(r)} H_0(r), e^{u(r)} H_2(r), r^2 K(r), r^2 \sin^2 \theta K(r)] \,.
\end{align}
We further write the scalar field perturbation as
\begin{align} \label{eq:scalar_perturbation}
\delta \Phi(t,r,\theta,\varphi) = \sum_{{\rm m}=-2}^{2} e^{-i\omega t} \frac{\phi_1(r)}{r} Y_{2 {\rm m}}(\theta,\varphi)\,.
\end{align}
By requiring $\delta \Phi$ to have the same time dependence as the background field, we assure that the perturbed energy-momentum tensor is static to linear order. 

We then insert Eqs.~(\ref{eq:metric_perturbation}) and (\ref{eq:scalar_perturbation}) into the linearized Einstein's equations, $\delta G^\alpha_{\,\beta} = 8\pi \delta T^\alpha_{\,\beta}$. The $\theta\theta$- minus the $\phi\phi$-component reveal that $H_2(r) = -H_0(r)$. Also, the $r\theta$-component can be used to relate $K'$ to $\phi_1$, $H_0$, and its derivative:
\begin{equation}
K' = - H_0' - H_0 v' - \frac{32\pi \phi_1 \phi'_0}{r}\,.
\end{equation}
Finally, the $rr$- minus the $tt$-component can be written as a master equation for $H_0$, with no further $K$ dependence:
\begin{align}
& H_0''(r) +\frac{4-r u'+r v'}{2 r} H_0'(r) \nonumber \\
&+\left (32\pi e^{u-v} \omega^2 \phi_0^2 - \frac{6 e^u}{r^2} + \frac{2 v'}{r}-\frac{u'v'}{2} - \frac{v^{\prime 2}}{2} + v''\right ) H_0 \nonumber \\
& =\frac{16 \pi}{r^2} \left[ 2 r e^{u-v} \omega^2 \phi_0 - (4 - r u' - r v')\phi'_0-2 r \phi''_0 \right]\phi_1\,. \label{eq:H0}
\end{align}
Additionally, the perturbed scalar wave equation reads
\begin{align}
&\left[ \frac{d^2}{d r^2_*} \!+ \omega^2 \! - e^v \! \left (\! \frac{7 - e^{-u}}{r^2} \!+\! \mu^2 \!-\! 8\pi \mu^2 \phi^2_0 + 32\pi e^{-u} \phi^{\prime 2}_0 \!\right)\right ] \! \phi_1 \nonumber \\
&= \left[ 2 r \omega^2 \phi_0 - e^v r \left( \mu^2 \phi_0 - e^{-u} v' \phi'_0 \right) \right] H_0\,, \label{eq:phi1}
\end{align}
where $r_*$ is defined through $dr_* = e^{(u-v)/2} dr$. Our task is to solve the coupled system formed by Eqs.~(\ref{eq:H0}) and (\ref{eq:phi1}) subject to suitable boundary conditions. Note that, again, it is possible to scale away the $\mu$ dependence of these equations by suitable variable redefinitions. 

Around $r=0$, the leading-order behaviour of $H_0(r)$ and $\phi_1(r)$ is given by
\begin{equation}
	H_0(r) \approx H_0^{(2)} r^2 + O(r^4), \quad
	\phi_1(r) \approx \phi_1^{(3)} r^3 + O(r^5).
\end{equation}
Since Eqs.~(\ref{eq:H0}) and (\ref{eq:phi1}) are invariant under an overall normalization of $H_0$ and $\phi_1$, we can fix, say, $H_0^{(2)}=1$ and retrieve the correct normalization a posteriori (in our case, this would be determined by the strength of the tidal environment). 
The value of $\phi_1^{(3)}$ is determined by imposing 
\begin{equation} \label{eq:BCphi1}
\lim_{r\to \infty} \phi_1(r) = 0.
\end{equation}
The procedure for constructing a numerical solution is entirely analogous to what was described previously: by fine-tuning the value of $\phi_1^{(3)}$, we guarantee that condition (\ref{eq:BCphi1}) is satisfied in a sufficiently large domain. 
Notice from Eq.~(\ref{eq:phi1}) that since $\phi_0(r)$ goes to zero exponentially as $r\to \infty$, the boundary condition (\ref{eq:BCphi1}) is not inconsistent with a polynomial growth for $H_0(r)$ in this limit, which is to be expected by Eq.~(\ref{eq:asymp_gtt}). 

For distances much larger than the typical size $R$ of the boson star, Eq.~(\ref{eq:H0}) takes the form
\begin{equation}\label{eq:Heq_asymp}
H''_0 + H'_0 \left( \frac{2}{r} + e^u \frac{2M}{r^2} \right) - H_0 \left( \frac{6 e^u}{r^2} + e^{2u} \frac{4 M^2}{r^4} \right)=0\,,
\end{equation}
where we used that $u'=-v'=-2M e^{u}/r^2$ in this regime.
Equation (\ref{eq:Heq_asymp}) has the general solution (in terms of associate Legendre functions):
\begin{equation}
H_0(r) \approx  c_1 Q_2^2(r/M - 1) + c_2 P_2^2(r/M-1),
\end{equation}
valid for $r\gg R$, and which behaves as
\begin{equation}
H_0(r) \approx 3 \left( \frac{r}{M}\right)^2 c_2 + O (r^3)+ \frac{8}{5} \left( \frac{M}{r}\right)^3 c_1 + O(r^{-4})
\end{equation}
as $r \to \infty$.
The growing piece of the metric perturbation accounts for the presence of an external tidal field, while the decaying piece encodes the response of the body. By matching this asymptotic form with Eq.~(\ref{eq:asymp_gtt}), one infers that $\lambda = (8 M^5/5) c_1/c_2$. Therefore, the value of $\lambda$ can be inferred from the numerical solution at $r = R_\textrm{ext} \gg R$ by evaluating $y := R_\textrm{ext} H'_0(R_\textrm{ext})/H_0(R_\textrm{ext})$ and computing \cite{Hinderer:2008}
\begin{align}\label{eq:lambda}
\lambda &= \frac{16}{15} M^5 (1-2 \mathcal{C})^2 [2 + 2\mathcal{C} (y-1) - y] \nonumber \\
&\times \{ 3 (1-2\mathcal{C})^2 [2-y+2\mathcal{C}(y-1)] \log(1-2\mathcal{C})
\nonumber \\
& + 2 \mathcal{C} [ 6 - 3 y + 3\mathcal{C} (5y - 8)] + 4\mathcal{C}^3 [13 - 11y 
\nonumber \\
& + \mathcal{C}(3y - 2) + 2 \mathcal{C}^2 (1+y)] \}^{-1},
\end{align}
where $\mathcal{C} := M / R_\textrm{ext}$. For sufficiently large extraction radius $R_\textrm{ext}$, the value thus obtained for $\lambda$ is independent of $R_\textrm{ext}$.

\begin{figure}[bt]
\includegraphics[width=8.4cm]{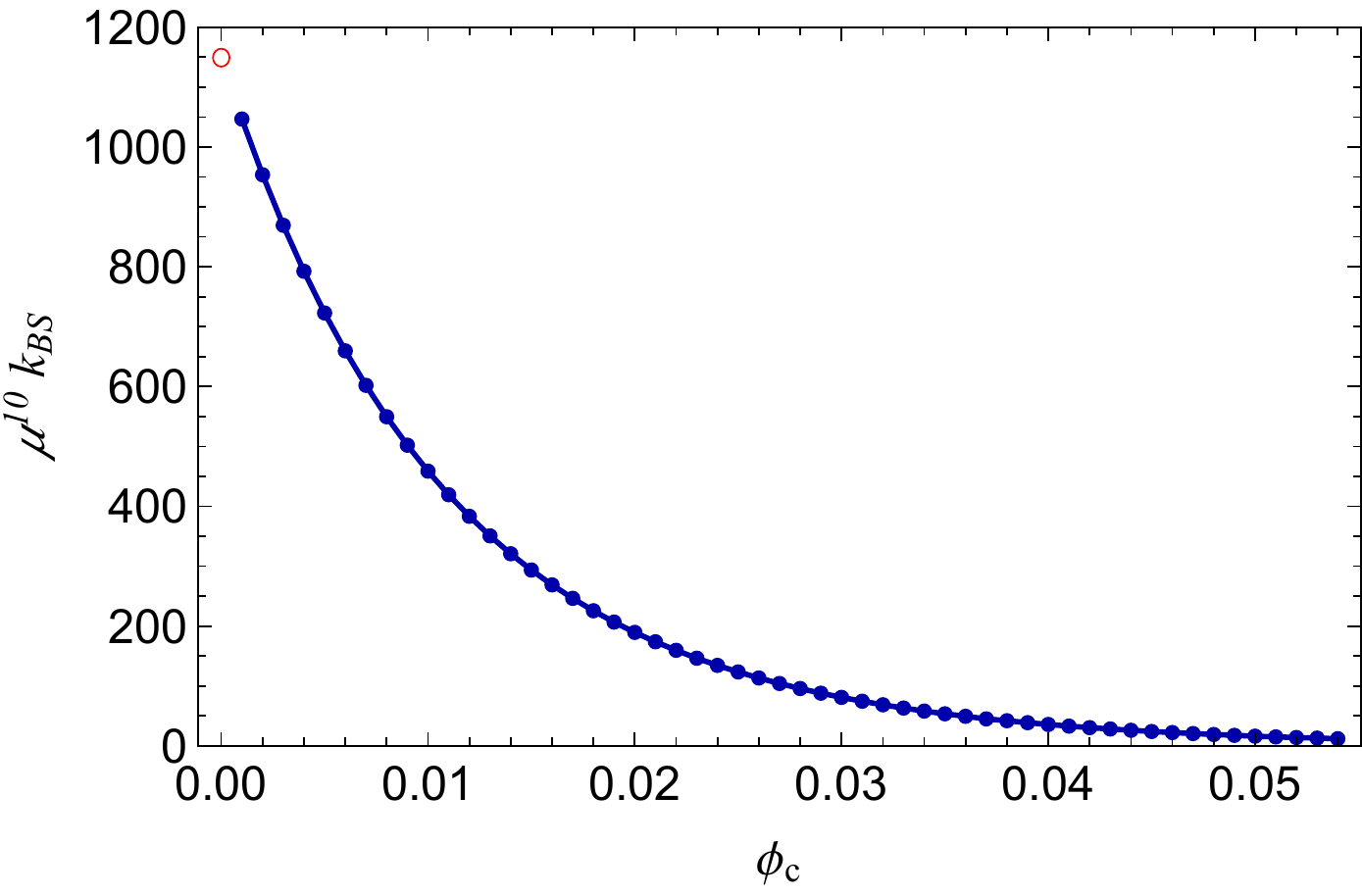}
\caption{For a sequence of equilibrium boson stars, characterized by the central value of the scalar field $\phi_c$, we plot the tidal number $k_{\rm BS} := \lambda M^5$, rescaled by $\mu^{10}$ to yield a dimensionless quantity. The Newtonian value, as computed in Sec.~\ref{sec:newtonian}, is shown as a red open circle.}
\label{fig:kBS}
\end{figure}

For fluid stars, the $l=2$ tidal Love number is usually defined as $k_2 := (3/2) R_s^{-5} \lambda$, where $R_s$ is the stellar radius. This is convenient not only because $k_2$ is a dimensionless quantity, but also because it has a meaningful Newtonian limit. 
Indeed, as we approach the Newtonian regime, the star becomes less compact and the stellar radius grows for a fixed mass. Since $\lambda$ typically grows as $R_s^5$, this definition guarantees that $k_2$ is finite in the limit of $R_s \to \infty$. 
A similar behaviour is found for boson stars as $\phi_c \to 0$ (or $M \to 0$; $R \to \infty$). In this limit, we can verify that $\lambda$ grows as $R^5$, as $M^{-5}$, or as $\phi_c^{-5/2}$. Since the usual definition of $k_2$ is quite sensitive to the value of the radius (as it enters to the 5th power), which is inherently ambiguous for boson stars, we choose instead to define the quantity 
\begin{equation} \label{eq:kBS}
k_{\rm BS} := \lambda M^5.
\end{equation}
This quantity is plotted as a function of $\phi_c$ in Fig.~\ref{fig:kBS}, and has a finite limit as $\phi_c \to 0$.

In the next subsection, we shall compute independently the value of $k_{\rm BS}$ in the Newtonian limit. While the calculation is mostly analogous to what we did so far, it offers a few new insights and works as a consistency check of the relativistic calculation. The Newtonian value of $k_{\rm BS}$ is exhibited as a red open circle in Fig.~\ref{fig:kBS} and is consistent with the trend of the relativistic curve.

\subsection{Newtonian limit}
\label{sec:newtonian}

In this subsection we repeat the steps of the relativistic calculation presented above, but for the Sch\"odinger-Poisson system, which follows as the Newtonian limit of the Einstein-Klein-Gordon equations (\ref{eq:EE})-(\ref{eq:KG}). For the sake of clarity, throughout this subsection we recover the lost factors of $c$'s and $G$'s.

\begin{figure}[tb]
\includegraphics[width=8.2cm]{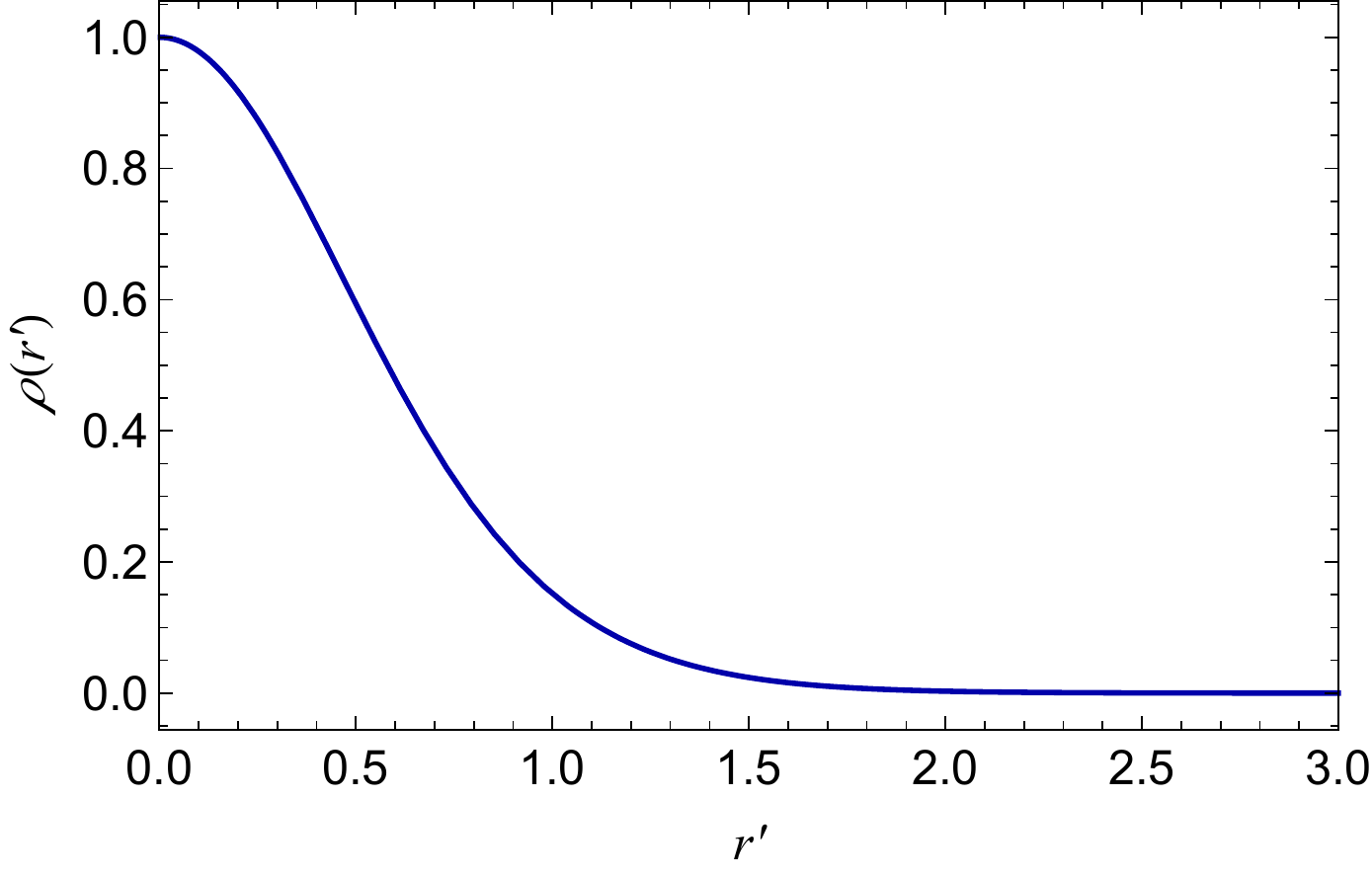}
\caption{Density profile rescaled by central density, $\rho(r) \equiv |\Psi(t,r)|^2/|\Psi(t,0)|^2$, as a function of $r' \equiv r |\Psi(t,0)|^{1/2}$, with $r$ measured in units of $\hbar/mc$. The rescaled energy parameter is given by $E'\equiv E/|\Psi(t,0)| = -3.256$, with $E$ in units of $mc^2$.}
\label{fig:density}
\end{figure}

The Newtonian limit of the Einstein-Klein-Gordon equations can be obtained by setting (in Cartesian coordinates)
\begin{align}
g_{tt} &= -c^2 [ 1 + 2U / c^2 + \mathcal{O}(c^{-4})] \, 
\nonumber \\
g_{ij} &= (1 - 2U/c^2 ) \delta_{ij} + \mathcal{O}(c^{-4}).
\end{align}
We further write $\Phi = e^{i m c^2 t/\hbar} \Psi$, thereby factoring out the high-frequency oscillations of the field. 
To leading order (in $1/c$), Eq.~(\ref{eq:EE}) gives rise to the Poisson equation, 
\begin{equation} \label{eq:poisson}
\nabla^2 U = 8\pi G (m/\hbar)^2 |\Psi|^2,
\end{equation}
while Eq.~(\ref{eq:KG}) reduces to the Schr\"{o}dinger equation 
\begin{equation} \label{eq:schr}
i\hbar \frac{\partial \Psi}{\partial t} = -\frac{\hbar^2}{2m} \nabla^2 \Psi + m U \Psi,
\end{equation}
where we used that $\mu := mc/\hbar$.
The Newtonian analogue of a boson star is then a stationary solution of the form $\Psi(t,r) = e^{i E t/\hbar} \psi_0(r)$, which can be constructed numerically along the same lines described in Sec.~\ref{sec:background}, and is depicted in Fig.~\ref{fig:density}.

Interestingly, Eqs.~(\ref{eq:poisson}) and (\ref{eq:schr}) are invariant under the transformation
\begin{equation} \label{eq:rescaling}
\Psi \to C \Psi, \quad U \to C U, \quad r \to C^{-1/2} r,  \quad t \to C^{-1} t,
\end{equation}
where $C$ is an arbitrary constant. 
This implies that, contrary to the case of boson stars, in which there is a family of solutions parametrized by the central value of the scalar field, stationary solutions of the Sch\"{o}dinger-Poisson equations with different values of $\psi_c = \psi_0(0)$ can be trivially related by a rescaling of the form (\ref{eq:rescaling}). In particular, the solution's total mass, defined as $M:= \lim_{r \to \infty} r U(r) / G$, transforms as $M \to C^5 M$ under a change of normalization $\Psi \to C \Psi$. 

In order to determine the tidal deformability of a ``Newtonian boson star'', we consider general pertubations of the form
\begin{equation}
U = U_0 + \delta U, \qquad \Psi = \Psi_0 + \delta \Psi,
\end{equation} 
where $U_0$ and $\Psi_0$ denote a background stationary solution and we write 
\begin{align*}
\delta U & = \frac{1}{2} \sum_{{\rm m}=-2}^2 e^{i Et/\hbar} H(r) Y_{2{\rm m}}(\theta,\varphi), \\
\delta \Psi &= \sum_{{\rm m}=-2}^2 e^{i Et/\hbar} \frac{\psi_1(r)}{r} Y_{2{\rm m}}(\theta,\varphi).
\end{align*}
To linear order, the perturbation equations become
\begin{equation} \label{eq:perturbN1}
H'' + \frac{2}{r} H' -\frac{6H}{r^2} = \frac{16 \pi G m^2}{\hbar^2 r} (\psi_1^*\psi_0 + \psi_1 \psi_0^*)
\end{equation}
and
\begin{equation} \label{eq:perturbN2}
\psi''_1 - \left [ \frac{6}{r^2} + \frac{2 m}{\hbar^2} (m U_0 + E) \right] \psi_1 = \frac{m^2}{\hbar^2} r H \psi_0.
\end{equation}
Given a background solution for $\psi_0$, $U_0$ and $E$, the perturbation equations can be solved numerically by a procedure entirely analogous to the one described in Sec.~\ref{sec:perturbations}: by adjusting the value of $\psi_1^{(3)} = \psi'''(0)$ one ensures that the boundary condition $\lim_{r\to \infty} \psi_1 (r) = 0$ is satisfied in a sufficiently large numerical domain.

For distances much larger than the typical size $R$ of the star, the source term in Eq.~(\ref{eq:perturbN1}) vanishes and $H(r)$ behaves as
\begin{equation}
H(r) \approx \frac{d_1}{r^3} + d_2 r^2.
\end{equation}
It is straightforward to infer that $\lambda = (1/3)d_1/d_2$ [see Eq.~(\ref{eq:asymp_gtt})]. Therefore, $\lambda$ can be computed from numerical data at an extraction radius $R_\textrm{ext} \gg R$ by 
\begin{equation}\label{eq:lambdaN}
\lambda = R_\textrm{ext}^5 \frac{2-y}{3(3 + y)},
\end{equation}
where again $y:=R_\textrm{ext} H'(R_\textrm{ext})/H(R_\textrm{ext})$. Equation (\ref{eq:lambdaN}) is the Newtonian limit of Eq.~(\ref{eq:lambda}), obtained by neglecting corrections of higher order in $\mathcal{C} = GM/(c^2 R_\textrm{ext})$. Again, for sufficiently large $R_\textrm{ext}$, the value thus obtained for $\lambda$ is independent of $R_\textrm{ext}$; indeed, as $R_\textrm{ext} \to \infty$, we have that $y \to 2$ in such a way that Eq.~(\ref{eq:lambdaN}) has a well-defined limit.

As discussed before, the Schr\"{o}dinger-Poisson equations remain invariant under the transformation (\ref{eq:rescaling}). 
However, the same is \textit{not} true for $\lambda$, which  transforms as $\lambda \to C^{-5/2} \lambda$. A more meaningful quantity is given, e.g., by $k_{\rm BS}^{(N)} = \lambda (G M/c^2)^5$, defined in analogy with Eq.~(\ref{eq:kBS}). This tidal number is not affected by a renormalization of the wave function and can be computed to be
\begin{equation} \label{eq:kBSN}
k_{\rm BS}^{(N)} \approx 1149 \times ( \hbar/mc )^{10}.
\end{equation}
As shown in Fig.~\ref{fig:kBS}, this value is consistent with the appropriate limit of the relativistic calculation.

%%%%%%%%%%%%%%%%%%%%%%%%%%%%%%%%%%%%%%%%%%%%%
\section{Wave description of collisionless dark matter}\label{sec3}

We shall now change gears from wave to particle dark matter models, in order to discuss an approximate mapping between them. 

\subsection{Basic definitions}\label{secwig}

A collisionless dark matter clump with total mass $M$ and particle mass $m$ is described by a phase-space density distribution $P({\bf x},{\bf p})$ obeying the collisionless Boltzmann (or Vlasov) equation,
\begin{align}\label{eqboltz}
\frac{\partial P}{\partial t} +\frac{\bf p}{m} \cdot \nabla_{\bf x} P - m \nabla_{\bf x} U \cdot \nabla_{\bf p} P=0\,,
\end{align}
where the gravitational potential $U({\bf x})$ must satisfy the Poisson equation
\begin{align}
\nabla_{\bf x}^2 U = 4 \pi m \int d^3 {\bf p} P({\bf x}, {\bf p})\,.
\end{align}
In addition, the phase-space density should be normalized with respect to the total particle number\footnote{Notice that, as long as the particle number is conserved, the concepts of phase-space density and probability density of a single particle are interchangeable, as they are proportional to each other. In other words, the phase-space density can be viewed as a probability density with total probability $M/m$. We shall not try to distinguish them in later analyses. The same considerations also apply for discussion below on the Wigner function, where we implicitly expand its definition to be the incoherent summation of the Wigner functions of $M/m$ particles.}
\begin{align} \label{eq:normP}
\int d^3 {\bf x} \int d^3 {\bf p}\, P({\bf x},{\bf p}) = \frac{M}{m}\,.
\end{align}

In order to introduce the mapping between cold dark matter clumps and Newtonian boson stars, let us first recall the concept of the Wigner function of a quantum mechanical system (for a more thorough introduction, we point e.g.~to Refs.~\cite{walls2007quantum,Scully1997}). For simplicity, we use the notation as in $1$-D systems, but it is straightforward to generalize the discussion to higher dimensions. The Wigner function is usually defined in terms of the density matrix $\hat \rho$ of the quantum system as \cite{Scully1997}
\begin{align}
W(x,p) := \frac{1}{\pi \hbar} \int \rho(x+y,x-y) e^{-2 i p y/\hbar} dy\,,
\end{align}
where $\rho(x_1,x_2) := \langle x_1|\hat{\rho}|x_2\rangle$. Correspondingly,
\begin{align}\label{eqrhow}
\rho(x_1,x_2) = \int W\left ( \frac{x_1+x_2}{2},p \right) e^{-i p (x_2-x_1)/\hbar} dp\,.
\end{align}

The Wigner function $W(x,p)$ satisfies many properties similar to the phase-space probability distribution $P(x,p)$. For example, by integrating out one of the variables ($x$ or $p$), it becomes the marginal probability distribution for the other variable. The main differences are that (i) the Wigner function can be negative (hence it is referred to as a \textit{quasiprobability} distribution), (ii) it obeys a different evolution equation than the classical Boltzmann equation. 

Specifically, from the evolution equation for the density matrix,
\begin{equation} \label{eq:master_equation}
\frac{\partial \hat \rho}{\partial t} = \frac{1}{i\hbar} [\hat H,\hat \rho],
\end{equation}
where $\hat H$ is the Hamiltonian and square brackets indicate the commutator, it follows that the time evolution of the Wigner function is governed by the Moyal equation:
\begin{align} \label{eq:moyal_equation}
\frac{\partial W(x,p)}{\partial t}+ \{W(x,p),\hat{H}(x,p)\}=0,
\end{align}
where $\{,\}$ is called the Moyal bracket:
\begin{align}
\{f,g\} := \frac{2}{\hbar} f(x,p) \sin \left[ \frac{\hbar
}{2}(\overleftarrow{\partial_x}\,\overrightarrow{ \partial_p} -\overleftarrow{\partial_p} \,\overrightarrow{\partial_x)}\right ] g(x,p).
\end{align}
Here $\overleftarrow{\partial}$ denotes the derivative operator acting on the terms to its left, and similarly for $\overrightarrow{ \partial}$. 

In particular, for a quadratic potential $U \propto x^2$ (as in the dominant, quadrupolar piece of a tidal field), the Moyal equation reduces to
\begin{align}
\partial_t W + \frac{p}{m} \partial_x W - m\, \partial_x U \partial_p W =0\,,
\end{align}
which is precisely the collisionless Boltzmann equation.
In general, after Taylor-expanding the sinusoidal operator, the Moyal evolution equation can be written as
\begin{align}\label{eq:evolution_equation}
&\partial_t W +\frac{p}{m} \partial_x W - m\, \partial_x U \, \partial_p W \nonumber \\
&+m \sum^\infty_{n=1} \frac{\hbar^{2n}}{(2n+1)!(2i)^{2n}}\partial^{2n+1}_p W \cdot\partial^{2n+1}_x U=0\,.
\end{align}
Note that, by taking the $\hbar \to 0$ limit in Eq.~(\ref{eq:evolution_equation}), and assuming $W$ is sufficiently well-behaved in this limit, it is possible to minimize the contributions from the terms in the second line, which embody the departure from the collisionless Boltzmann equation.

\subsection{The particle-wave correspondence}

The similarity between the Moyal equation in the $\hbar \to 0$ limit and the collisionless Boltzmann equation is at the core of the proposals of Refs.~\cite{widrow1993using,schaller2014new,Davies1997}, which trade the evolution of the phase-space probability distribution $P({\bf x},{\bf p})$ by the evolution of a suitable quantum-mechanical system. We describe the basic idea in the following, illustrating the principles using a system with one spatial and one momentum dimension, as previously. 

Given an initial phase-space density distribution $P(x,p)$, which we identify with the Wigner function, we define an effective density matrix $\sigma(x_1,x_2)$ in analogy with Eq.~(\ref{eqrhow}):
\begin{align}\label{eqsig1}
\sigma(x_1,x_2) =\int P\left ( \frac{x_1+x_2}{2},p \right) e^{-i p (x_2-x_1)/\varpi} dp\,
\end{align}
and, conversely, 
\begin{align}\label{eqsig2}
P(x,p) =\int \frac{1}{\pi \varpi} \sigma(x+y, x-y)e^{-2 i p y/\varpi} dy\,,
\end{align}
where $\varpi$ is a yet undetermined constant that plays the role of $\hbar$ in Eq.~(\ref{eqrhow}). 

The next step is to find a representation of $\sigma(x_1,x_2)$ in terms of a set of wave functions $\{\psi_i(x)\}$ as
\begin{align}\label{eqdec}
\sigma(x_1,x_2) = \sum_i p_i \psi_i(x_1) \psi_i^\dagger(x_2) \,,
\end{align}
with $\sum_i p_i=1$, and $p_i \ge 0$. Note that it is always possible to decompose the density matrix into a convex combination of pure states as above (e.g. using the Schmidt decomposition). In general, however, this procedure is not unique, which means one has the freedom to choose the state basis to, e.g., minimize the complexity (such as the number of states) in the summation.

With a decomposition as in Eq.~(\ref{eqdec}), the master equation (\ref{eq:master_equation}) for a self-gravitating system can be written as a set of coupled Schr\"{o}dinger-Poisson equations,
\begin{subequations}\label{eqm}
\begin{align}
& i \hbar \, \partial_t \psi_i +\frac{\hbar^2}{2 m} \nabla^2_x \psi_i - m U \psi_i=0\,, \\
& \nabla_x^2 U = 4 \pi M \sum_i p_i |\psi_i|^2\,,
\end{align}
\end{subequations}
where we have modified the Poisson equation so as to adopt the standard normalization 
\begin{align}
\int d x |\psi_i|^2= 1\,,
\end{align}
while the information on the particle number or total mass of the field is encoded in the coefficient $M$. 

Therefore, after decomposing the original phase-space density distribution into a set of wave functions through the steps (\ref{eqsig1}) and (\ref{eqdec}), the time evolution of the system is performed through the Schr\"{o}dinger-Poisson equations (\ref{eqm}), instead of using the collisionless Boltzmann equation (\ref{eqboltz}). Finally, one recovers a phase-space description by applying the inverse mapping (\ref{eqsig2}). In \cite{widrow1993using, Davies1997}, the Husimi quasiprobability distribution was used instead of the Wigner function as a phase-space representation of the wave functions, but the underlying idea is the same. 
By following this method, we trade solving the Boltzmann equation for $P({\bf x},{\bf p})$, which in general is six-dimensional, by solving three-dimensional coupled Schr\"{o}dinger-Poisson equations for a set of complex wave functions $\psi_i({\bf x})$. This is potentially computationally advantageous, as long as the set of wave functions is not too large. 

If we are to follow the procedure outlined above, there are two issues that need to be clarified. First, we need to appropriately specify the quantity $\varpi$ that plays the role of $\hbar$ in Eq.~(\ref{eqsig1}), such that the effective density matrix $\sigma$ has a simple wave decomposition as in Eq.~(\ref{eqdec}). Second, we need to clarify the systematic error introduced by using the Moyal equation, instead of the Boltzmann equation, to evolve $P$. We discuss these issues in the following.

%%%%%%%%%%%%%%%%%%%%%%%%%
\subsubsection{Specifying $\varpi$} \label{sec:identification}

In general, $\sigma(x_1,x_2)$, as given in Eq.~(\ref{eqsig1}), describes a ``mixed state" and can have multiple decompositions as in Eq.~(\ref{eqdec}).
General, ``brute-force'', algorithms to perform this decomposition exist, as described, for instance, in Ref.~\cite{schaller2014new}, but they are hardly optimal. 
The task of finding the optimal decomposition, which makes use of the lowest number of wave functions but is still accurate, is a nontrivial one, and must be done in a case by case basis. 
Here we do not attempt at providing a general prescription for the decomposition of arbitrary functions $P$ (see Ref.~\cite{schaller2014new} for that). Instead, we illustrate how this decomposition can be performed in a few examples, emphasizing how a suitable choice of $\varpi$ can enable a description of the state in terms of a small set of wave functions (although occasionally at the expense of a loss in accuracy).

As a first example, consider the case where $P$ is initially Gaussian, 
\begin{equation} \label{eq:Pgaussian}
P(x,p) = \frac{1}{\sqrt{2 \pi} \sigma_x}\frac{1}{\sqrt{2 \pi} \sigma_p} \exp\left[-\frac{x^2}{2 \sigma^2_x} -\frac{p^2}{2 \sigma^2_p}\right]\,,
\end{equation}
where $\sigma_x$ and $\sigma_p$ are the variances in the position and momentum distributions, respectively. By Eq.~(\ref{eqsig1}), the corresponding effective density matrix is 
\begin{equation*}
\sigma(x_1,x_2) = \frac{1}{\sqrt{2 \pi} \sigma_x} \exp\left[-\frac{(x_1+x_2)^2}{8\sigma_x^2} - \frac{(x_1-x_2)^2 \sigma_p^2}{2\varpi^2}\right].
\end{equation*}
In order to decompose $\sigma(x_1,x_2)$ as in Eq.~(\ref{eqdec}), one needs to specify the value of $\varpi$, that here plays the role of $\hbar$. In the example above, a particularly simple decomposition ensues if one sets $\varpi = 2 \sigma_x \sigma_p$. In this case, the density matrix describes a pure state: $\sigma(x_1,x_2) = \psi(x_1)\psi^*(x_2)$, where
\begin{equation*}
\psi(x)=\left (\frac{1}{\sqrt{2 \pi }\sigma_x}\right )^{1/2}   \exp\left[-\frac{x^2}{4\sigma_x^2}\right] \,.
\end{equation*}
Therefore, in some cases, such as when $P$ is approximately Gaussian, a choice for $\varpi$  arises naturally, namely, $\varpi \sim \sigma_x \sigma_p$.
Another such case is when the $P$ function is reconstructed from stationary solutions of the Schr\"{o}dinger-Poisson equations (the Newtonian analogue of boson stars), with a mass-density distribution as in Fig.~\ref{fig:density}. These can be seen as approximations to stationary solutions of the collisionless Boltzmann equation and will be our model for dark matter clumps when we translate our results from Sec.~\ref{sec2}.

Another simple case is when one does not have access to, or is not interested in, the entire phase-space information contained in $P$. For instance, one may want to specify the mass distribution alone, which is encoded in the diagonal components of the effective density matrix. In this case, in principle a pure state decomposition of $\sigma$ is possible, by setting $\sigma ({\bf x}_1,{\bf x}_2) = \psi({\bf x}_1) \psi^*({\bf x}_2)$, with
\begin{equation}\label{eq:NFW}
\psi({\bf x}) = \sqrt{\bar{\rho}({\bf x})} e^{i \phi({\bf x})},
\end{equation}
where $\bar{\rho}({\bf x})$ is the specified mass density profile and $\phi({\bf x})$ is an arbitrary phase that can be set as desired. For instance, one could want to reproduce the Navarro-Frenk-White profile for dark matter halos, $\bar{\rho}(r) = \rho_s/[(r/R_s)(1+r/R_s)^2]$, where $\rho_s$ and $R_s$ are the characteristic halo density and radius, or some smoothed version of that profile.
Notice that, in this construction, there is still some freedom in the momentum distribution, as one can adjust the value of $\varpi$, as well as specify the phase $\phi({\bf x})$. For instance, if one wants to match the variance $\sigma_{\bf p}$ of some specified momentum distribution, typically this can be accomplished by choosing $\varpi$ of the order of $\sigma_{\bf p} \sigma_{\bf x}$ (together with a phase $\phi({\bf x})$ that does not vary rapidly in the angular directions), where $\sigma_{\bf x}$ gives the spatial variation scale of $\bar{\rho}({\bf x})$. (However, in order to match the momentum distribution instead of its variance, a more complex decomposition seems to be inevitable, with an ensemble of phases $\phi_i(x)$ with higher spatial frequencies.)

The discussions above suggest that a simple reconstruction of the probability distribution, using a small number of wave functions, can often be achieved if $\varpi \sim \sigma_x \sigma_p$. 
However, this choice comes at the expense of losing the freedom to choose $\varpi$ small and thereby minimize the error terms in the evolution equation (\ref{eq:evolution_equation}). Conversely, if $\varpi$ is chosen to be $\ll \sigma_x \sigma_p$ as in \cite{widrow1993using,schaller2014new}, the error in the evolution equation would usually be smaller, but many short-wavelength wave functions would typically be needed to reconstruct $\sigma$. In this case, the number of wave functions to be evolved would be larger and the method would lose some of its computational advantage.
In general, the choice of $\varpi$ depends on the accuracy required in a given problem or timescale, and the available state decomposition of a particular $P$.

In this work we shall focus on the case where $\varpi \sim \sigma_x \sigma_p$, which means that the effective particle de Broglie wavelength $\lambda_\textrm{deB} := \varpi/mv$ (with $v\sim |(\partial\psi/\partial t)/\nabla\psi|$) is comparable to the spatial variation scale of the phase-space distribution. In this case, we may achieve lower accuracy in describing the system's evolution, so it would be better if the disagreement with the exact evolution was small within dynamical timescales. Let us now turn to this point.

\subsubsection{Validation}

With the mappings (\ref{eqsig1}) and (\ref{eqsig2}) between the phase-space density $P$ and the effective density matrix $\sigma$, we can discuss the error involved in using the wave approximation above, in terms of evolution equations for $\sigma$. On the one hand, as $P$ must follow the collisionless Boltzmann equation (\ref{eqboltz}), consequently $\sigma$ must follow
\begin{align}\label{eqexact}
\frac{\partial \sigma}{\partial t} + \frac{i}{\varpi} & \left [ m(x_2-x_1) \cdot \nabla U \left ( \frac{x_1+x_2}{2}\right ) \sigma \right. \nonumber \\
& \left. -\frac{ \varpi^2 }{2 m}\left ( \frac{\partial^2 \sigma}{\partial x^2_2}-\frac{\partial^2 \sigma}{\partial x^2_1}\right ) \right ] =0\,.
\end{align}
On the other hand, by using the wave approximation above, we are effectively evolving $\sigma$ using the master equation (\ref{eq:master_equation}), which can be written in coordinate representation as
\begin{equation}\label{eqapprox}
\frac{\partial \sigma}{\partial t} +\frac{i}{\varpi} \left [  m[U(x_2)-U(x_1)]\sigma-\frac{\varpi^2}{2 m}\left ( \frac{\partial^2 \sigma}{\partial x^2_2}-\frac{\partial^2 \sigma}{\partial x^2_1}\right ) \right ] =0\,,
\end{equation}
with $\varpi$ again playing the role of $\hbar$. In the expressions above, if $U(x)$ is the gravitational potential generated by the system itself, it satisfies
\begin{align}
\nabla^2_x U = 4 \pi m \sigma(x,x)\,.
\end{align}

Comparing Eqs.~(\ref{eqexact}) and (\ref{eqapprox}), we find that the only difference is the potential related term, where the $(x_2-x_1)\cdot \nabla U$ term in Eq.~(\ref{eqexact}) can be viewed as a first order Taylor expansion of $U(x_2)-U(x_1)$, in Eq.~(\ref{eqapprox}), around the middle point $(x_2+x_1)/2$. If the potential is linear or quadratic in $x$, which is the case for tidal fields, such difference vanishes and the two equations are identical. In general, the higher-order terms in the Taylor expansion will cause systematic errors when we use Eq.~(\ref{eqapprox}) to evolve the system. Obviously the error should be smaller for near-diagonal terms in the density matrix, where $|x_1-x_2|$ is small. In fact, when we take the $\varpi \rightarrow 0$ limit, we see from Eq.~(\ref{eqsig1}) that the effective density matrix $\sigma$ associated with a given $P$ becomes increasingly diagonal: the integral is non-negligible only when $x_1 \approx x_2$, due to the rapid phase oscillations in the integrand otherwise. Thus, the influence of error terms becomes less important in this regime, in agreement with our previous discussion. 
Nevertheless, if we are mainly interested in the evolution of the spatial mass distribution $\sigma(x,x)$, Eq.~(\ref{eqapprox}) actually provides a reasonable description within the gravitational dynamical timescale, even not in the $\varpi \rightarrow 0$ limit. This is illustrated in Appendix \ref{apnum} using a numerical simulation. 

%%%%%%%%%%%%%%%%%%%%%%%%%%%%%%%%%%%%%%%%%%%%%

\section{Mapping to particle dark matter clumps} \label{sec4}

When Widrow and Kaiser introduced their wave-mechanical approach for cold dark matter in \cite{widrow1993using}, their main interest was to explore the potential computational advantages of this method in cosmological simulations, since one would replace the task of solving the six-dimensional Boltzmann equation by a set of three-dimensional Schr\"{o}dinger-Poisson equations. In order to ensure accuracy of the approximation over cosmological timescales, they had to perform a decomposition of the phase-space density distribution into multiple pure states; but even with this overhead in the number of differential equations to be solved, the method was shown to have a comparable computational cost to the more standard $N$-body simulations \cite{widrow1993using,Davies1997}.

In this work we envision a somewhat different application of the wave-mechanical approach described above, namely, as a tool to investigate dynamical properties of dense dark matter clumps (DMCs) under gravity. The systems we are interested in are clumps of sub-parsec scales, which are allowed by many dark matter models as a result of primordial adiabatic isocurvature fluctuations \cite{PhysRevD.50.769}, cosmological phase transitions \cite{PhysRevD.59.043517}, topological defects \cite{silk1992dmc}, accretion on primordial black holes \cite{bertschinger1985self}, etc. DMCs with with larger densities are likely to survive from tidal interactions with the galactic environment (e.g., see discussion in \cite{PhysRevLett.117.141102,PhysRevD.73.063504,Berezinsky2013}), but the viable spectra of their mass, density and radius distribution is still highly uncertain. 

We are particularly interested in the possibility of probing the existence and properties of these DMCs through the emission coming from their tidal interactions with the surrounding medium. Indeed, dense DMCs could possibly form binaries with stars or other dense DMCs within the globular cluster (via three body interaction or dynamical scattering), they could be captured by other star binaries, and they may pass the vicinity of supermassive black holes and give rise to tidal disruption events  \cite{PhysRevD.93.043508}. In addition to the gamma rays that could be generated by dark matter annihilation within these DMCs, these tidal encounters could also generate gravitational waves, with frequency lying in the band of advanced ground-based interferometers \cite{LIGOreport} and/or future space-based interferometers \cite{amaro2012elisa}.

The detectability of gravitational waves (GWs) from dark matter clumps would of course depend heavily on their properties, most notably their compactness and abundance, of which we are still uncertain. 
One possibility would be to observe GWs from binaries consisting of a compact dark matter clump and another compact body. If there exist in nature dark matter clumps with compactness $GM/(Rc^2)\gtrsim 0.1$, i.e., comparable with neutron stars, GWs emitted during tidal encounters could be measured by ground-based detectors, such as LIGO. If the density of dark mater clumps is close to white dwarfs, there is still a chance they can be detected by instruments such as LISA or DECIGO. Of course, if dark matter only agglutinates in low density configurations, there is little chance of observing these binaries in the gravitational wave window.

Another possibility would be to look at environmental effects of dark matter clumps acting on GW emitting systems within it. For example, a compact binary inside a dark matter clump would feel the dynamical friction due to the gravitational interaction with dark matter particles. The detailed strength of this dynamical friction force could be different from the one for a gaseous medium with similar density, but they should have a similar order of magnitude. Therefore by observing the long-time evolution of the gravitational waveform, it could be possible to tell whether there is such an additional frictional force (see Ref.~\cite{macedo2013into} for a study in a related set-up).

The wave-mechanical approach described in Sec.~\ref{sec3} seems particularly suitable to deal with systems undergoing tidal interactions. In particular, if we neglect self-gravity, the potential is dominated by the quadratic, quadrupolar piece of the tidal field. In this case, it is straightforward to see (cf.~Appendix A) that a wave-mechanical approach agrees exactly with previous results using standard methods for the evolution of the system \cite{PhysRevD.93.043508}. In the presence of self-gravity, the potential is no longer purely quadratic, but as argued above and illustrated in the numerical simulations presented in Appendix B, it is still possible to give a wave-mechanical description to the dynamical evolution of the system which is both simple and accurate within the relevant dynamical timescale. By simplicity we mean that we can work in the regime where the effective particle wavelength is comparable to the spatial variation scale of the phase-space distribution. 
Although this approximation sacrifices some of the accuracy of the method (as discussed in Sec.~\ref{sec2}), it seems convenient for analytical and perturbative analyses, which help us build physical intuition into the system. 

Here we assume that the unperturbed density profile for the DMC is approximately given by a stationary solution of the Newton-Poisson equations, as in Fig.~\ref{fig:density}, and that the DMC is perturbed by an external quadrupolar tidal field. 
Given that the density distribution of DMCs is much less constrained than, e.g., dark matter halos, the assumption above seems reasonable, but it could also be lifted in favour of some more realistic model.
In the following we translate the results in Sec.~\ref{sec2} for the tidal deformability of Newtonian boson stars to an order of magnitude estimation of the corresponding quantity for dark matter clumps. We leave the task of making this mapping more precise to future work, but refer to the numerical analysis in Appendix B as evidence that the quantities computed here should offer reasonable approximations.

To proceed it is useful to rewrite the Schr\"{o}dinger-Poisson equations [cf.~Eq.~(\ref{eqm})] in terms of dimensionless (spatial, temporal and mass) variables, obtained through 
\begin{subequations}\label{eqmap}
\begin{align}
t \rightarrow & \left ( \frac{m c^2}{\hbar}\right )^{-1} \times \tilde{t}  \,, \\ 
{\bf x} \rightarrow & \left ( \frac{m c}{\hbar}\right)^{-1}  \times \tilde{\bf x} \,, \\
M \rightarrow & \left ( \frac{\hbar c}{G m }\right ) \times \tilde{M}\,,
\end{align}
\end{subequations}
in which case we can write
\begin{equation}\label{eqrescale}
i \partial_{\tilde{t}} \psi = -\frac{1}{2} \tilde{\nabla}^2 \psi - \tilde{M} \psi \int d^3\tilde{\bf x}' \frac{|\psi(\tilde{{\bf x}}')|^2}{|\tilde{{\bf x}} - \tilde{{\bf x}}'|} \,,
\end{equation}
where the wave function normalization is fixed to be
\begin{align}
\int d^3 \tilde{x}' |\psi({\bf x}')|^2=1.
\end{align}
Equations (\ref{eqmap}) can be used to map between truly wave-mechanical systems, such as boson stars, and cold dark matter clumps with a related mass density profile. 
For example, a DMC with total mass $M_{\rm DM}$, particle mass $m_{\rm DM}$, and radius $R_{\rm DM}$ can be mapped, within our approximation scheme, to a boson star with total mass $M_{\rm BS}$ and scalar field mass $m_{\rm BS}$, as long as 
\begin{align} \label{eq:map1}
\frac{M_{\rm DM} m_{\rm DM} }{\varpi} \approx \frac{M_{\rm BS} m_{\rm BS}}{\hbar}\,,
\end{align}
which follows from the third line of Eq.~\eqref{eqmap} once we require the same dimensionless mass $\tilde{M}$ in both problems. We must now make a choice for the constant $\varpi$: Following the discussion in Sec.~\ref{sec:identification}, we take 
\begin{equation}\label{eq:varpi_choice}
\varpi \sim \Delta x \Delta p \approx R_{\rm DM} m_{\rm DM}\sqrt{G M_{\rm DM}/R_{\rm DM}}\,,
\end{equation}
where $\Delta x$ and $\Delta p$ are the typical spatial and momentum scales of the system. Then Eq.~(\ref{eq:map1}) may be rewritten as
\begin{align}
\frac{M_{\rm DM}}{M_{\rm BS}} &\sim 5.1 \times 10^{-8} \times \nonumber \\
& \left ( \frac{M_{\rm DM}}{ M_{\odot}}\right )^{1/2} \left ( \frac{R_{\rm DM}}{10^6 R_{\odot}}\right )^{1/2} \left ( \frac{m_{\rm BS} c^2}{10^{-23} {\rm eV}}\right )\,.
\end{align}

With the mapping established between a DMC and a boson star, we can then discuss the relation between their perturbative properties. 
The tidal deformability $\lambda$, for example, can be obtained by first computing the rescaled (dimensionless) quantity $\tilde{\lambda}:= (mc/\hbar)^5 \lambda$, and then restoring the physical dimensions to obtain the final value. This indicates that
\begin{align}
\lambda_{\rm DM} \approx \lambda_{\rm BS} \left ( \frac{m_{\rm DM} \hbar }{m_{\rm BS} \varpi }\right)^{-5} \approx \lambda_{\rm BS} \left ( \frac{M_{\rm DM}}{M_{\rm BS}}\right )^5\,.
\end{align}
In particular, for dark matter clumps, the tidal Love number $k_{\rm DM} := \lambda (GM_{\rm DM}/c^2)^5$, defined in analogy with Eq.~(\ref{eq:kBS}), is given by [cf.~Eq.~(\ref{eq:kBSN})]
\begin{equation}
k_{DM} \approx 1149 \times \left( \frac{\varpi}{m_{\rm DM} c} \right)^{10}
\sim 1149 \times \left( \frac{GM_{\rm DM} R_{\rm DM}}{c^2} \right)^5 
\end{equation}
where again we used Eq.~(\ref{eq:varpi_choice}).

Other perturbative properties, such as the quasinormal mode frequencies of oscillations, allow for analogous mappings. We can first evaluate the rescaled mode frequency for the rescaled equation, Eq.~(\ref{eqrescale}), and then scale it back to original configurations. At the end, the mode frequencies are (approximately) related by 
\begin{align}
\omega_{\rm DM} \approx \omega_{\rm BS} \left ( \frac{m_{\rm DM} \hbar }{m_{\rm BS} \varpi }\right) \approx \omega_{\rm BS} \left ( \frac{M_{\rm BS}}{M_{\rm DM}}\right )\,,
\end{align}
where we used Eqs.~(\ref{eqmap}) and (\ref{eq:map1}).
Thus, results obtained for the quasinormal modes of boson stars \cite{Yoshida1994,PhysRevD.88.064046} can be translated into order-of-magnitude estimates for cold dark matter clumps.

As a final comment, we notice that after a stationary boson star configuration is mapped to a DMC, it generally does not remain stationary, as the underlying evolution equation becomes the collisionless Boltzmann equation instead of the Moyal equation. However, due to the close proximity between these two equations, as discussed in Sec.~\ref{sec3}, it is possible that the DMC just oscillates around a new equilibrium configuration which is not far from its initial state. In this case, the oscillation frequency and tidal deformability of the new equilibrium configuration should be roughly consistent with the numbers we obtain by analyzing the perturbations of the original boson star. One can understand this intuitively by considering the case of a simple quantum harmonic oscillator. If we add to the potential a perturbation of order $\mathcal{\epsilon}$, the original stationary, ground-state wavefunction will not be the ground state of the new system (not even an equilibrium state in general). However, it is reasonable to expect, through eigenvalue perturbation theory, that the eigenfrequencies are only perturbed by order $\mathcal{O}(\epsilon)$ as well (except for the special cases where the eigenfrequencies are degenerate). In our case, the perturbation scales as the third derivative of the potential [see discussion around Eq.~(\ref{eq:diff2})], but a concrete error estimate is more involved since we need to solve self-consistently for the potential, which introduces nonlinearities in the equations.
In the future, it would be interesting to test the claims above with further numerical experiments.

\section{Conclusion} \label{sec5}

Boson stars have been studied as dark matter candidates and possible exotic compact objects. Their existence and properties can be investigated e.g.~through the gravitational waves they would emit due to tidal encounters with other astrophysical bodies. An important quantity that can in principle be extracted from these waveforms are the tidal Love numbers, which reveal information about the internal structure of the boson star. In this work we computed the leading order response of a minimally coupled boson star under an external tidal field, which is encoded in its ($l=2$) tidal Love number. Our main results are summarized in Fig.~\ref{fig:kBS}.

As a nontrivial application of these results, in the second part of the work we discussed how to map them to particle dark matter clumps. 
This is based on trading the phase-space density distribution for the dark matter particles, which follows the collisionless Boltzmann equation, by a Wigner quasiprobability distribution following the Moyal equation. Equivalently, the problem can be formulated in terms of wave functions evolving according to a set of coupled Schr\"{o}dinger-Poisson equations, which reveals the connection with Newtonian boson stars.
In contrast to previous treatments, where many short-wavelength wave functions were required in order to reduce the error in approximating the Boltzmann equation, we work in the regime where the effective particle wavelength is comparable to the spatial variation scale of the phase-space distribution. We discuss the error incurred in adopting this procedure, and argue that it should be small within the relevant dynamical timescales. 

This (approximate) dual wave description seems specially convenient for analytical and perturbative analyses, and provides an interesting link to compare the phenomenology of cold dark matter models and scalar wave dark matter models, while their mass scales usually differ by many orders of magnitude. However, we emphasize that, in cases where the gravitational potential is highly non-smooth, i.e, the average variation scale is much smaller than the size of the DM clump, such approximation might be badly behaved. As the density distribution of DM clumps is currently poorly constrained by observations, this mapping should be treated as a conjecture which needs to be examined in different circumstances.

\acknowledgments
HY thanks Haixing Miao and Ue-Li Pen for very instructive discussions. We thank Eric Poisson for reading over a draft of this manuscript and giving many useful comments. R.M.~acknowledges financial support from Conselho Nacional de Desenvolvimento Cient\'{i}fico e Tecnol\'{o}gico (CNPq) and from a CITA National Fellowship to the University of Guelph. H.~Y.~acknowledges support from the Perimeter Institute for Theoretical Physics and the Institute for Quantum Computing. Research at Perimeter Institute is supported by the government of Canada and by the Province of Ontario through Ministry of Research and Innovation. 

\appendix
\section{Consistency check: tidal compression without self-gravity} \label{apcons}
 
In \cite{PhysRevD.93.043508}, a dark matter clump (DMC) under a strong tidal gravitational field was analyzed (see also \cite{stone2013consequences}). In particular, the self-gravity produced by the DMC was neglected and the motion of dark matter particles in each direction was treated as independent. As shown in \cite{PhysRevD.93.043508} and \cite{stone2013consequences},  given the initial phase-space density, its later distribution can be obtained by applying the conservation of phase-space density implied by Liouville's theorem and solving for the trajectory of each dark matter particle. We shall use this example to illustrate how to apply the effective wave description to derive later distributions. As the potential here is purely quadratic, the effective approach should produce the same result as the one by the Boltzmann equation (or Liouville's theorem). 
As the motion of particles is independent in the $x,y,z$ directions, we can simplify the problem to $1$-D, and take $U_{\rm tide} =\alpha x^2$ ($\alpha >0$, the case with $\alpha <0$ can be analyzed similarly).

Similarly to \cite{PhysRevD.93.043508}, we assume a Gaussian initial distribution, as given in Eq.~(\ref{eq:Pgaussian}).
Correspondingly, we can choose 
\begin{equation}
\psi(t=0,x)=\left (\frac{1}{\sqrt{2 \pi }\sigma_x}\right )^{1/2}   e^{-\frac{x^2}{4\sigma_x^2}} \,,
\end{equation}
and $\varpi =2 \sigma_x \sigma_p$. It is straightforward to check that the ``Wigner function" of this wave function reproduces the initial phase-space density in  Eq.~(\ref{eq:Pgaussian}).
The Schr\"{o}dinger equation we need to solve is
\begin{align}
 i \varpi \frac{\partial \psi}{\partial t} +\frac{\varpi^2}{2 m} \nabla^2_x \psi - m \alpha x^2 \psi=0\,.
\end{align}

Instead of trying to solve for $\psi(t)$ directly, we notice that a Gaussian wave packet remains Gaussian under a quadratic potential, which means we only need to know its mean and variance to determine its shape. To obtain mean and variance, we adopt the Heisenberg picture:
\begin{equation}\label{eqheisen}
 \dot {\hat{x}} = \frac{\hat{p}}{m}\,, \qquad
 \dot{\hat{p}} = -2 m \alpha \hat{x}\,.
\end{equation}
These equations have the following solution in terms of $\hat{x}_0, \hat{p}_0$:
\begin{align}
\hat{x} = \hat{x}_0 \cos \omega t+\frac{\hat{p}_0}{m \omega}\sin\omega t,\quad \hat{p}=\hat{p}_0 \cos \omega t - \omega m \hat{x}_0\sin\omega t,
\end{align}
with $\omega^2 = 2 \alpha$. Therefore the variance at time $t$ is just
\begin{align*}
{\rm Var}[\hat{x}] &=\langle \hat{x}^2 \rangle = {\rm Var}[\hat{x}_0] \cos^2\omega t+\frac{{\rm Var}[ \hat{p}_0]}{m^2\omega^2} \sin^2\omega t \nonumber \\
&= \sigma^2_x \cos^2\omega t+\frac{\sigma^2_p}{m^2\omega^2}  \sin^2\omega t\,,\nonumber \\
 {\rm Var}[\hat{p}]&= \langle \hat{p}^2 \rangle =\sigma^2_p \cos^2\omega t+m^2\omega^2 \sigma_x^2 \sin^2\omega t \,, \nonumber \\
{\rm Cov}[\hat{x} \hat{p}]&=\frac{1}{2} \langle \hat{x} \hat{p}+\hat{p} \hat{x}\rangle = \frac{\sin 2\omega t}{2}\left (\frac{\sigma^2_p}{m\omega}-m\omega \sigma_x^2 \right ),
\end{align*}
where we have used the fact that $\langle \hat{x}_0 \hat{p}_0 \rangle=0$. With these variances we can easily reconstruct the corresponding Wigner function, as it must retain its Gaussian shape. We can check that it is consistent with the calculation using Liouville's theorem in phase space in \cite{PhysRevD.93.043508}.  In fact, if we use Liouville's theorem and write down the equations of motion in phase space, they are just the ``classical" version of the Heisenberg equations of motion in Eq.~(\ref{eqheisen}).
In this sense, the equivalence of the two approaches is rather natural.

\section{Numerical comparison in a 1-D problem} \label{apnum}

In this Appendix, we compare the evolution of the effective density matrix $\sigma(x_1,x_2)$ according to Eq.~(\ref{eqexact}) and Eq.~(\ref{eqapprox}), which are associated to the collisionless Boltzmann equation and the Moyal equation, respectively. As we saw in Appendix A, in the approximation of no self-gravity, a DMC in a tidal environment can be described by either of these equations with identical results. A natural question that follows is whether an approximate agreement is maintained when self-gravity is taken into account. In this Appendix we address this question. However, in order to avoid the complications introduced by having to solve self-consistently for the gravitational potential, we assume instead an artificial attractive potential in additional to the tidal field; namely
\begin{equation} \label{eq:U}
U(x) = \alpha_1 x^2 - \alpha_2 e^{-x^2/4 a^2}.
\end{equation}
Throughout this Section $c$'s and $G$'s will be conveniently restored.
The constant $\alpha_1$ scales as $\alpha_1 \sim G M'/R'^3$, where $M'$ is the mass of a second body, responsible for generating the tidal field, and $R'$ is the distance of the DMC to it. The constant $\alpha_2$ scales as $\alpha_2 \sim G M/R$, where $M$ and $R$ are the mass and radius of the DMC. Naturally, $a \sim R$. The contributions of the two terms in Eq.~(\ref{eq:U}) are comparable if $R'$ is of the order of the \textit{tidal radius}, defined as $R_t = (M'/M)^{1/3} R$.

It is convenient to define the dimensionless quantities
\begin{equation}\label{eq:dimensionless_vars}
\tilde{t} := t \frac{mc^2}{\varpi}, \qquad \tilde{x} := x \frac{mc}{\varpi}, \qquad \tilde{U} := \frac{U}{c^2}.
\end{equation}
In terms of them Eqs.~(\ref{eqexact}) and (\ref{eqapprox}) assume the following forms:
\begin{equation} \label{eqexactb}
\frac{1}{i} \frac{\partial \sigma}{\partial \tilde{t}} + (\tilde{x}_2-\tilde{x}_1)\cdot \nabla \tilde{U}\left ( \frac{\tilde{x}_1+\tilde{x}_2}{2}\right ) \sigma - \frac{1}{2}\left ( \frac{\partial^2 \sigma}{\partial \tilde{x}^2_2}-\frac{\partial^2 \sigma}{\partial \tilde{x}^2_1}\right ) =0\,
\end{equation}
and
\begin{equation}\label{eqapproxb}
\frac{1}{i} \frac{\partial \sigma}{\partial \tilde{t}} + [\tilde{U}(\tilde{x}_2)-\tilde{U}(\tilde{x}_1)]\sigma - \frac{1}{2} \left ( \frac{\partial^2 \sigma}{\partial \tilde{x}^2_2}-\frac{\partial^2 \sigma}{\partial \tilde{x}^2_1}\right ) =0\,.
\end{equation}

At $t=0$, we consider two possible initial conditions for the density matrix $\sigma$. The first is a Gaussian profile:
\begin{equation}\label{eq:id}
\sigma^{(1)}(t=0;\tilde{x}_1,\tilde{x}_2) =  \frac{1}{\sqrt{2\pi} \sigma_x} e^{-(\tilde{x}_1^2 + \tilde{x}_2^2)/(4 \sigma_x^2)}.
\end{equation}
As we saw in Sec.~\ref{sec:identification}, this is the density matrix obtained from a Gaussian phase-space density distribution $P$ in the limit where the particle de Broglie wavelength is comparable to the variation scale of the particle distribution (i.e.~$\varpi \sim \sigma_x \sigma_p$). As a second possible initial condition, we set
\begin{equation} \label{eq:id2}
\sigma^{(2)}(t=0;\tilde{x}_1,\tilde{x}_2) =  \frac{N a^3}{x_1^2 x_2^2} (1-e^{-\tilde{x}_1^2/\sigma_x^2})(1-e^{-\tilde{x}_2^2/\sigma_x^2}),
\end{equation}
where $N \approx 0.51$ is a normalization factor. This density matrix has a power law, rather than exponential, spatial decay.
We then let the initial data (\ref{eq:id}) and (\ref{eq:id2}) evolve according to Eq.~(\ref{eqexactb}) and Eq.~(\ref{eqapproxb}), and compare the respective outcomes $\sigma^{(i)}_\textrm{exact}(t;x_1,x_2)$ and $\sigma^{(i)}_\textrm{approx}(t;x_1,x_2)$ ($i=1,2$) at later times. 

For the results presented below, we set $\sigma_x = 5$, $\alpha_1 = 10^{-4} (mc^2/\varpi)^2$, $\alpha_2 = 0.5 c^2$, and $a = 10 \varpi/mc$. From the scaling considerations below Eq.~(\ref{eq:U}), we can see that $R' \approx 3.7 R_t$ for this choice of parameters. Also, it is useful to define $T := \sqrt{R^3/GM}$, which gives the dynamical timescale of the DMC. Defining a dimensionless timescale as in Eq.~(\ref{eq:dimensionless_vars}), we find that $\tilde{T} := T mc^2/\varpi \approx 15$ for the parameters above. 

\begin{figure}[tb]
\includegraphics[width=8.4cm]{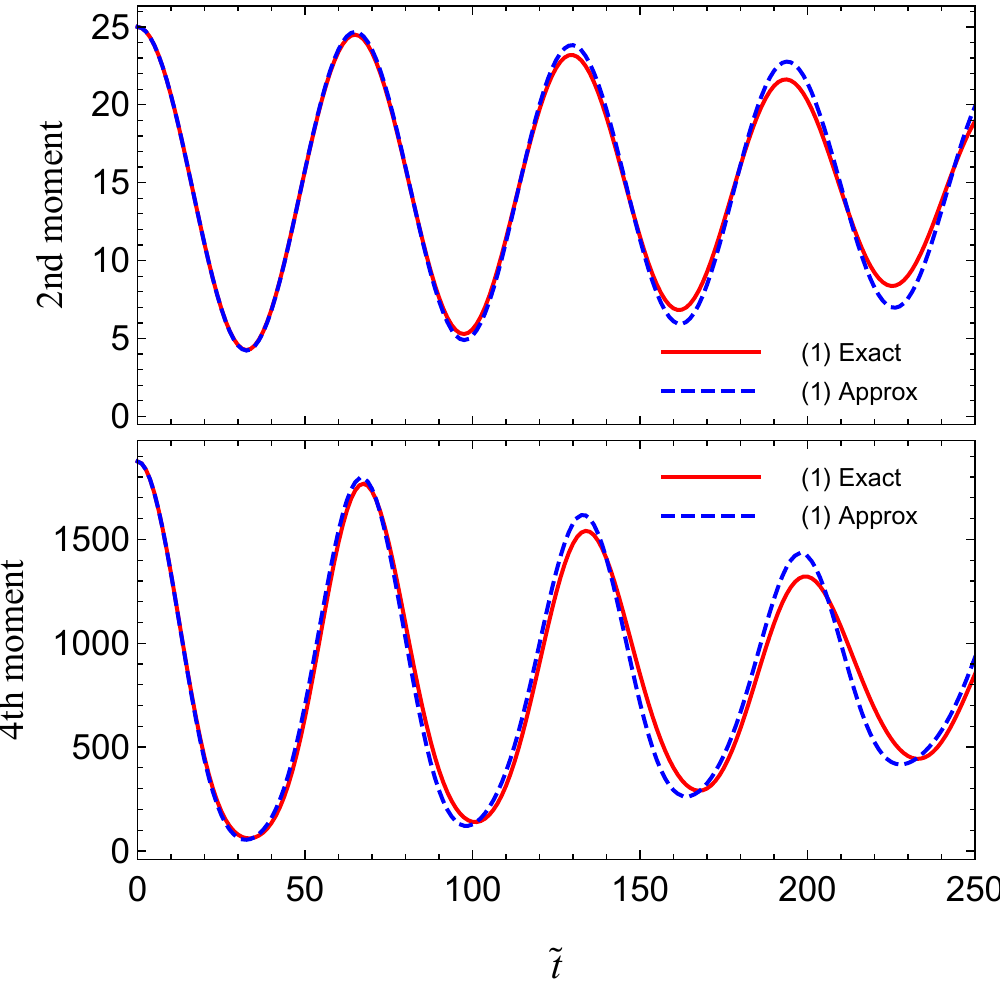}
\caption{Second and fourth moments of the density distributions $\sigma^{(1)}_\textrm{exact}(x,x)$ and $\sigma^{(1)}_\textrm{approx}(x,x)$ as functions of time.}
\label{fig:moments}
\end{figure}

\begin{figure}[tb]
\includegraphics[width=8.4cm]{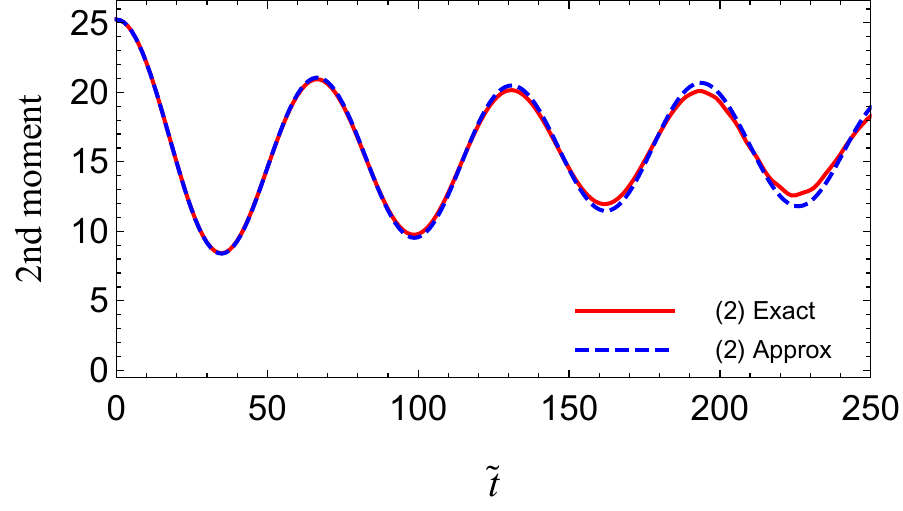}
\caption{Second moment of the density distributions $\sigma^{(2)}_\textrm{exact}(x,x)$ and $\sigma^{(2)}_\textrm{approx}(x,x)$ as functions of time. Note that since $\sigma^{(2)}(x,x)$ has a $x^{-4}$ fall off [see Eq.~(\ref{eq:id2})], it does not have a fourth or higher moments.}
\label{fig:moments2}
\end{figure}

The numerical setup is the following. We discretize Eqs.~(\ref{eqexactb}) and (\ref{eqapproxb}) in a uniform grid with $n^2$ nodes $(\tilde{x}_{1,i},\tilde{x}_{2,j}) = (-L/2 + i \Delta x, -L/2 + j \Delta x)$, where $i,j \in \{0,1, ..., n-1\}$, $\Delta x := L/(n-1)$, and $L$ determines the size of the numerical domain. For the results presented here, we take $L=200$ and $n=401$. The space derivatives are discretized with central finite differences and we evolve in time using a fourth-order Runge-Kutta algorithm. The time step is given by $\Delta t = \kappa (\Delta x)^2$, with $\kappa = 0.25$.

For many applications (such as computing the gravitational response), we are primarily interested in the density distribution $\sigma(x,x)$. In Fig~\ref{fig:moments}, we show the second and fourth moments of the density distributions $\sigma^{(1)}_\textrm{exact}(x,x)$ and $\sigma^{(1)}_\textrm{approx}(x,x)$, as functions of time. We see that the distributions are very close to each other within the dynamical timescale, and that they follow a similar evolution for an even longer period of time.
In Fig.~\ref{fig:moments2}, we show the second moment of the distributions $\sigma^{(2)}_\textrm{exact}(x,x)$ and $\sigma^{(2)}_\textrm{approx}(x,x)$, and we find an equally good agreement. 

\begin{figure}[htb]
\includegraphics[width=8.5cm]{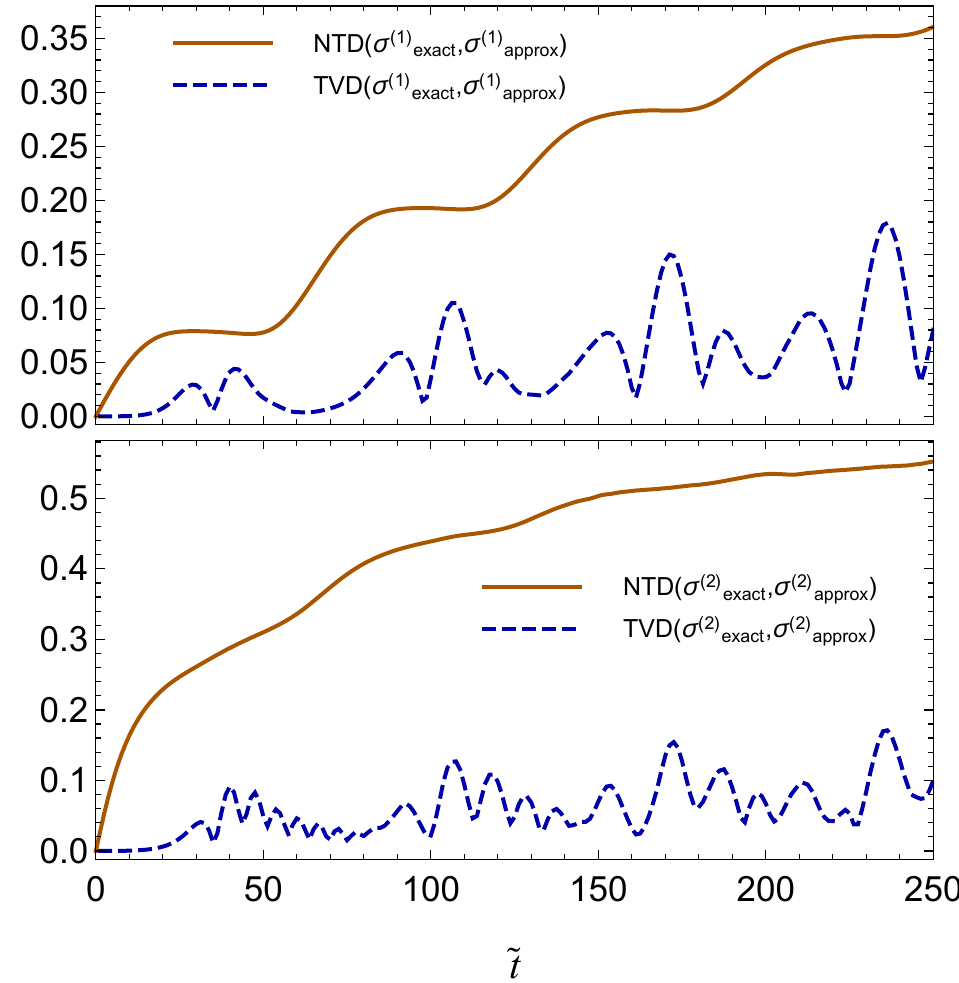}
\caption{Normalized trace distance and total variation distance between $\sigma^{(i)}_\textrm{exact}$ and $\sigma^{(i)}_\textrm{approx}$, as a function of time. We define ${\rm NTD}(\rho,\sigma) := ||\rho - \sigma||_1/(||\rho||_1+||\sigma||_1)$, where $||\rho||_1:=\frac{1}{2} \textrm{Tr} [\sqrt{\rho^\dagger \rho}]$, and ${\rm TVD}(\rho,\sigma):= \frac{1}{2} \sum_x | \rho(x,x) - \sigma(x,x) | \Delta x $. Note that the total variation distance is restricted to the diagonals of the matrices, which are valid probability distributions. 
Both distance measures lie in the interval $[0,1]$.
}
\label{fig:td}
\end{figure}

In Fig.~\ref{fig:td}, we show the (normalized) trace distance between the ``density matrices'' $\sigma^{(i)}_\textrm{exact}(x_1,x_2)$ and $\sigma^{(i)}_\textrm{approx}(x_1,x_2)$, and the total variation distance between the distributions $\sigma^{(i)}_\textrm{exact}(x,x)$ and $\sigma^{(i)}_\textrm{approx}(x,x)$. The plot indicates, in agreement with our previous discussion, that these matrices differ mostly in their nondiagonal components.
Moreover, while the distance between the {\it full} density matrices is larger for the evolution of the initial condition (\ref{eq:id2}) than for (\ref{eq:id}), the distance between their {\it diagonal} part is comparable in both cases.

One could attempt to estimate the timescale at which the error matrix $\delta \sigma := \sigma_\textrm{exact}(x_1,x_2) - \sigma_\textrm{approx}(x_1,x_2)$ becomes significant by considering its evolution equation,
\begin{align}\label{eq:diff}
&\frac{1}{i} \frac{\partial \delta \sigma}{\partial \tilde{t}} + [\tilde{U}(\tilde{x}_2)-\tilde{U}(\tilde{x}_1)]\delta \sigma - \frac{1}{2} \left ( \frac{\partial^2 \delta\sigma}{\partial \tilde{x}^2_2}-\frac{\partial^2 \delta \sigma}{\partial \tilde{x}^2_1}\right ) \,\nonumber \\
& \approx \frac{1}{24} (\tilde{x}_2-\tilde{x}_1)^3 \tilde{U}''' \left(\frac{\tilde{x}_1+\tilde{x}_2}{2}\right) \sigma_\textrm{exact}\,,
\end{align}
where we have Taylor-expanded the potential term and kept only the leading contribution. By assumption, $\delta \sigma$ and its spatial derivatives vanish at $t=0$. Thus, at early times, the evolution of the error matrix is driven by the inhomogeneous term in the right-hand side of Eq.~(\ref{eq:diff}), and one could estimate a timescale $\Delta t$ from
\begin{equation}\label{eq:diff2}
| \delta \sigma | \approx \frac{\Delta t}{24} \left| (\tilde{x}_2-\tilde{x}_1)^3 \tilde{U}''' \left(\frac{\tilde{x}_1+\tilde{x}_2}{2}\right) \sigma_\textrm{exact} \right|.
\end{equation}
Computing the quantities in the right-hand side at $t=0$, for the potential given in Eq.~(\ref{eq:U}) and the initial data (\ref{eq:id}), and estimating the norm $||\delta \sigma|| := \sqrt{\int{\delta \sigma^* \delta \sigma dx_1 dx_2}}$, one obtains $||\delta \sigma||/||\sigma_\textrm{exact}|| \approx \Delta t / 150$. A similar estimate is obtained by averaging Eq.~(\ref{eq:diff2}) in time.
This would indicate that a typical fractional error of 10\% would occur after $\Delta t \approx 15$. This estimation is roughly consistent with the values obtained for the trace distance (see Fig.~\ref{fig:td}), which depends on the full density matrices, but clearly overestimates the error in the diagonal components of $\sigma$ (see Figs.~\ref{fig:moments} and \ref{fig:td}). The reason is that the evolution of $\delta \sigma (x,x)$ does not respond to the driving term in Eq.~(\ref{eq:diff}) directly, but only via the diffusive terms that we are implicitly neglecting. As a result, in our numerical simulations we see that the error in the diagonal part of the density matrix accumulates much more slowly than what our naive estimation would suggest.

Our numerical experiments support our claim that, even in the limit where the effective particle de Broglie wavelength is not tuned to be small, but is taken to be comparable with the size of system, the error incurred in the wave-mechanical method described here is small at least within the dynamical timescale of the system. Moreover, the error in the density matrix is mainly felt by its nondiagonal components. Note that these inaccurate far-off-diagonal components in $\sigma$ would mainly affect the low-momentum part of the phase space if we use Eq.~(\ref{eqsig2}) to reconstruct the phase-space distribution, as the $e^{2 i p y/\varpi}$ factor becomes highly oscillatory when the magnitude of $p, y$ is large. This is consistent with the observation that the mass distribution is less affected by the systematic error, as the change of mass density in short timescales is more vulnerable to ``fast" particles. As a last comment we note that the $P$ function obtained from evolving $\sigma$ according to Eq.~(\ref{eqapprox}) is essentially the Wigner function of Sec.~\ref{secwig}, which may become negative at part of the phase space during evolution. In fact, in most cases the negative value region is mainly located at the ``low-momentum" part of the phase space. Again, an accurate description of the mass distribution already reveals a lot of information about the system, especially its gravitational properties.

\bibliography{References}

\begin{thebibliography}{44}
\expandafter\ifx\csname natexlab\endcsname\relax\def\natexlab#1{#1}\fi
\expandafter\ifx\csname bibnamefont\endcsname\relax
  \def\bibnamefont#1{#1}\fi
\expandafter\ifx\csname bibfnamefont\endcsname\relax
  \def\bibfnamefont#1{#1}\fi
\expandafter\ifx\csname citenamefont\endcsname\relax
  \def\citenamefont#1{#1}\fi
\expandafter\ifx\csname url\endcsname\relax
  \def\url#1{\texttt{#1}}\fi
\expandafter\ifx\csname urlprefix\endcsname\relax\def\urlprefix{URL }\fi
\providecommand{\bibinfo}[2]{#2}
\providecommand{\eprint}[2][]{\url{#2}}

\bibitem[{\citenamefont{Jetzer}(1992)}]{Jetzer1992}
\bibinfo{author}{\bibfnamefont{P.}~\bibnamefont{Jetzer}},
  \bibinfo{journal}{Phys. Rep.} \textbf{\bibinfo{volume}{220}},
  \bibinfo{pages}{163} (\bibinfo{year}{1992}).

\bibitem[{\citenamefont{Schunck and Mielke}(2003)}]{Schunck2003}
\bibinfo{author}{\bibfnamefont{F.~E.} \bibnamefont{Schunck}} \bibnamefont{and}
  \bibinfo{author}{\bibfnamefont{E.~W.} \bibnamefont{Mielke}},
  \bibinfo{journal}{Classical and Quantum Gravity}
  \textbf{\bibinfo{volume}{20}}, \bibinfo{pages}{R301} (\bibinfo{year}{2003}).

\bibitem[{\citenamefont{Liebling and Palenzuela}(2012)}]{Liebling2012}
\bibinfo{author}{\bibfnamefont{S.~L.} \bibnamefont{Liebling}} \bibnamefont{and}
  \bibinfo{author}{\bibfnamefont{C.}~\bibnamefont{Palenzuela}},
  \bibinfo{journal}{Living Rev. Relativity} \textbf{\bibinfo{volume}{15}},
  \bibinfo{pages}{6} (\bibinfo{year}{2012}).

\bibitem[{\citenamefont{Cardoso et~al.}(2016)\citenamefont{Cardoso, Franzin,
  and Pani}}]{PhysRevLett.116.171101}
\bibinfo{author}{\bibfnamefont{V.}~\bibnamefont{Cardoso}},
  \bibinfo{author}{\bibfnamefont{E.}~\bibnamefont{Franzin}}, \bibnamefont{and}
  \bibinfo{author}{\bibfnamefont{P.}~\bibnamefont{Pani}},
  \bibinfo{journal}{Phys. Rev. Lett.} \textbf{\bibinfo{volume}{116}},
  \bibinfo{pages}{171101} (\bibinfo{year}{2016}).

\bibitem[{\citenamefont{Sin}(1994)}]{Sin1994}
\bibinfo{author}{\bibfnamefont{S.-J.} \bibnamefont{Sin}},
  \bibinfo{journal}{Phys. Rev. D} \textbf{\bibinfo{volume}{50}},
  \bibinfo{pages}{3650} (\bibinfo{year}{1994}).

\bibitem[{\citenamefont{Lee and Koh}(1996)}]{Lee1996}
\bibinfo{author}{\bibfnamefont{J.-w.} \bibnamefont{Lee}} \bibnamefont{and}
  \bibinfo{author}{\bibfnamefont{I.-g.} \bibnamefont{Koh}},
  \bibinfo{journal}{Phys. Rev. D} \textbf{\bibinfo{volume}{53}},
  \bibinfo{pages}{2236} (\bibinfo{year}{1996}).

\bibitem[{\citenamefont{Detweiler}(1980)}]{PhysRevD.22.2323}
\bibinfo{author}{\bibfnamefont{S.}~\bibnamefont{Detweiler}},
  \bibinfo{journal}{Phys. Rev. D} \textbf{\bibinfo{volume}{22}},
  \bibinfo{pages}{2323} (\bibinfo{year}{1980}).

\bibitem[{\citenamefont{Choptuik and Pretorius}(2010)}]{PhysRevLett.104.111101}
\bibinfo{author}{\bibfnamefont{M.~W.} \bibnamefont{Choptuik}} \bibnamefont{and}
  \bibinfo{author}{\bibfnamefont{F.}~\bibnamefont{Pretorius}},
  \bibinfo{journal}{Phys. Rev. Lett.} \textbf{\bibinfo{volume}{104}},
  \bibinfo{pages}{111101} (\bibinfo{year}{2010}).

\bibitem[{\citenamefont{Cardoso et~al.}(2017)\citenamefont{Cardoso, Franzin,
  Maselli, Pani, and Raposo}}]{cardoso2017testing}
\bibinfo{author}{\bibfnamefont{V.}~\bibnamefont{Cardoso}},
  \bibinfo{author}{\bibfnamefont{E.}~\bibnamefont{Franzin}},
  \bibinfo{author}{\bibfnamefont{A.}~\bibnamefont{Maselli}},
  \bibinfo{author}{\bibfnamefont{P.}~\bibnamefont{Pani}}, \bibnamefont{and}
  \bibinfo{author}{\bibfnamefont{G.}~\bibnamefont{Raposo}},
  \bibinfo{journal}{arXiv preprint arXiv:1701.01116}  (\bibinfo{year}{2017}).

\bibitem[{\citenamefont{Sennett et~al.}(2017)\citenamefont{Sennett, Hinderer,
  Steinhoff, Buonanno, and Ossokine}}]{sennett2017}
\bibinfo{author}{\bibfnamefont{N.}~\bibnamefont{Sennett}},
  \bibinfo{author}{\bibfnamefont{T.}~\bibnamefont{Hinderer}},
  \bibinfo{author}{\bibfnamefont{J.}~\bibnamefont{Steinhoff}},
  \bibinfo{author}{\bibfnamefont{A.}~\bibnamefont{Buonanno}}, \bibnamefont{and}
  \bibinfo{author}{\bibfnamefont{S.}~\bibnamefont{Ossokine}},
  \bibinfo{journal}{arXiv preprint arXiv:1704.08651}  (\bibinfo{year}{2017}).

\bibitem[{\citenamefont{Zwicky}(1933)}]{Zwicky1933}
\bibinfo{author}{\bibfnamefont{F.}~\bibnamefont{Zwicky}},
  \bibinfo{journal}{Helvetica Physica Acta} \textbf{\bibinfo{volume}{6}},
  \bibinfo{pages}{110} (\bibinfo{year}{1933}).

\bibitem[{\citenamefont{Steigman et~al.}(2012)\citenamefont{Steigman, Dasgupta,
  and Beacom}}]{PhysRevD.86.023506}
\bibinfo{author}{\bibfnamefont{G.}~\bibnamefont{Steigman}},
  \bibinfo{author}{\bibfnamefont{B.}~\bibnamefont{Dasgupta}}, \bibnamefont{and}
  \bibinfo{author}{\bibfnamefont{J.~F.} \bibnamefont{Beacom}},
  \bibinfo{journal}{Phys. Rev. D} \textbf{\bibinfo{volume}{86}},
  \bibinfo{pages}{023506} (\bibinfo{year}{2012}).

\bibitem[{\citenamefont{Hu et~al.}(2000)\citenamefont{Hu, Barkana, and
  Gruzinov}}]{Hu2000}
\bibinfo{author}{\bibfnamefont{W.}~\bibnamefont{Hu}},
  \bibinfo{author}{\bibfnamefont{R.}~\bibnamefont{Barkana}}, \bibnamefont{and}
  \bibinfo{author}{\bibfnamefont{A.}~\bibnamefont{Gruzinov}},
  \bibinfo{journal}{Phys. Rev. Lett.} \textbf{\bibinfo{volume}{85}},
  \bibinfo{pages}{1158} (\bibinfo{year}{2000}).

\bibitem[{\citenamefont{Akerib et~al.}(2017)\citenamefont{Akerib, Alsum,
  Ara\'ujo, Bai, Bailey, Balajthy, Beltrame, Bernard, Bernstein, Biesiadzinski
  et~al.}}]{LUX}
\bibinfo{author}{\bibfnamefont{D.~S.} \bibnamefont{Akerib}},
  \bibinfo{author}{\bibfnamefont{S.}~\bibnamefont{Alsum}},
  \bibinfo{author}{\bibfnamefont{H.~M.} \bibnamefont{Ara\'ujo}},
  \bibinfo{author}{\bibfnamefont{X.}~\bibnamefont{Bai}},
  \bibinfo{author}{\bibfnamefont{A.~J.} \bibnamefont{Bailey}},
  \bibinfo{author}{\bibfnamefont{J.}~\bibnamefont{Balajthy}},
  \bibinfo{author}{\bibfnamefont{P.}~\bibnamefont{Beltrame}},
  \bibinfo{author}{\bibfnamefont{E.~P.} \bibnamefont{Bernard}},
  \bibinfo{author}{\bibfnamefont{A.}~\bibnamefont{Bernstein}},
  \bibinfo{author}{\bibfnamefont{T.~P.} \bibnamefont{Biesiadzinski}},
  \bibnamefont{et~al.} (\bibinfo{collaboration}{LUX Collaboration}),
  \bibinfo{journal}{Phys. Rev. Lett.} \textbf{\bibinfo{volume}{118}},
  \bibinfo{pages}{021303} (\bibinfo{year}{2017}).

\bibitem[{\citenamefont{Tan et~al.}(2016)\citenamefont{Tan, Xiao, Cui, Chen,
  Chen, Fang, Fu, Giboni, Giuliani, Gong et~al.}}]{PandaXII}
\bibinfo{author}{\bibfnamefont{A.}~\bibnamefont{Tan}},
  \bibinfo{author}{\bibfnamefont{M.}~\bibnamefont{Xiao}},
  \bibinfo{author}{\bibfnamefont{X.}~\bibnamefont{Cui}},
  \bibinfo{author}{\bibfnamefont{X.}~\bibnamefont{Chen}},
  \bibinfo{author}{\bibfnamefont{Y.}~\bibnamefont{Chen}},
  \bibinfo{author}{\bibfnamefont{D.}~\bibnamefont{Fang}},
  \bibinfo{author}{\bibfnamefont{C.}~\bibnamefont{Fu}},
  \bibinfo{author}{\bibfnamefont{K.}~\bibnamefont{Giboni}},
  \bibinfo{author}{\bibfnamefont{F.}~\bibnamefont{Giuliani}},
  \bibinfo{author}{\bibfnamefont{H.}~\bibnamefont{Gong}}, \bibnamefont{et~al.}
  (\bibinfo{collaboration}{PandaX-II Collaboration}), \bibinfo{journal}{Phys.
  Rev. Lett.} \textbf{\bibinfo{volume}{117}}, \bibinfo{pages}{121303}
  (\bibinfo{year}{2016}).

\bibitem[{\citenamefont{Aprile et~al.}(2016)\citenamefont{Aprile, Aalbers,
  Agostini, Alfonsi, Amaro, Anthony, Arneodo, Barrow, Baudis, Bauermeister
  et~al.}}]{Xenon100}
\bibinfo{author}{\bibfnamefont{E.}~\bibnamefont{Aprile}},
  \bibinfo{author}{\bibfnamefont{J.}~\bibnamefont{Aalbers}},
  \bibinfo{author}{\bibfnamefont{F.}~\bibnamefont{Agostini}},
  \bibinfo{author}{\bibfnamefont{M.}~\bibnamefont{Alfonsi}},
  \bibinfo{author}{\bibfnamefont{F.~D.} \bibnamefont{Amaro}},
  \bibinfo{author}{\bibfnamefont{M.}~\bibnamefont{Anthony}},
  \bibinfo{author}{\bibfnamefont{F.}~\bibnamefont{Arneodo}},
  \bibinfo{author}{\bibfnamefont{P.}~\bibnamefont{Barrow}},
  \bibinfo{author}{\bibfnamefont{L.}~\bibnamefont{Baudis}},
  \bibinfo{author}{\bibfnamefont{B.}~\bibnamefont{Bauermeister}},
  \bibnamefont{et~al.} (\bibinfo{collaboration}{XENON Collaboration}),
  \bibinfo{journal}{Phys. Rev. D} \textbf{\bibinfo{volume}{94}},
  \bibinfo{pages}{122001} (\bibinfo{year}{2016}).

\bibitem[{\citenamefont{Ackermann et~al.}(2015)\citenamefont{Ackermann, Albert,
  Anderson, Atwood, Baldini, Barbiellini, Bastieri, Bechtol, Bellazzini,
  Bissaldi et~al.}}]{PhysRevLett.115.231301}
\bibinfo{author}{\bibfnamefont{M.}~\bibnamefont{Ackermann}},
  \bibinfo{author}{\bibfnamefont{A.}~\bibnamefont{Albert}},
  \bibinfo{author}{\bibfnamefont{B.}~\bibnamefont{Anderson}},
  \bibinfo{author}{\bibfnamefont{W.~B.} \bibnamefont{Atwood}},
  \bibinfo{author}{\bibfnamefont{L.}~\bibnamefont{Baldini}},
  \bibinfo{author}{\bibfnamefont{G.}~\bibnamefont{Barbiellini}},
  \bibinfo{author}{\bibfnamefont{D.}~\bibnamefont{Bastieri}},
  \bibinfo{author}{\bibfnamefont{K.}~\bibnamefont{Bechtol}},
  \bibinfo{author}{\bibfnamefont{R.}~\bibnamefont{Bellazzini}},
  \bibinfo{author}{\bibfnamefont{E.}~\bibnamefont{Bissaldi}},
  \bibnamefont{et~al.} (\bibinfo{collaboration}{The Fermi-LAT Collaboration}),
  \bibinfo{journal}{Phys. Rev. Lett.} \textbf{\bibinfo{volume}{115}},
  \bibinfo{pages}{231301} (\bibinfo{year}{2015}).

\bibitem[{\citenamefont{Gaskins}(2016)}]{Gaskins:2016}
\bibinfo{author}{\bibfnamefont{J.~M.} \bibnamefont{Gaskins}},
  \bibinfo{journal}{Contemporary Physics} \textbf{\bibinfo{volume}{57}},
  \bibinfo{pages}{496} (\bibinfo{year}{2016}).

\bibitem[{\citenamefont{Abbott et~al.}(2016{\natexlab{a}})\citenamefont{Abbott,
  Abbott, Abbott, Abernathy, Acernese, Ackley, Adams, Adams, Addesso, Adhikari
  et~al.}}]{PhysRevLett.116.061102}
\bibinfo{author}{\bibfnamefont{B.~P.} \bibnamefont{Abbott}},
  \bibinfo{author}{\bibfnamefont{R.}~\bibnamefont{Abbott}},
  \bibinfo{author}{\bibfnamefont{T.~D.} \bibnamefont{Abbott}},
  \bibinfo{author}{\bibfnamefont{M.~R.} \bibnamefont{Abernathy}},
  \bibinfo{author}{\bibfnamefont{F.}~\bibnamefont{Acernese}},
  \bibinfo{author}{\bibfnamefont{K.}~\bibnamefont{Ackley}},
  \bibinfo{author}{\bibfnamefont{C.}~\bibnamefont{Adams}},
  \bibinfo{author}{\bibfnamefont{T.}~\bibnamefont{Adams}},
  \bibinfo{author}{\bibfnamefont{P.}~\bibnamefont{Addesso}},
  \bibinfo{author}{\bibfnamefont{R.~X.} \bibnamefont{Adhikari}},
  \bibnamefont{et~al.} (\bibinfo{collaboration}{LIGO Scientific Collaboration
  and Virgo Collaboration}), \bibinfo{journal}{Phys. Rev. Lett.}
  \textbf{\bibinfo{volume}{116}}, \bibinfo{pages}{061102}
  (\bibinfo{year}{2016}{\natexlab{a}}).

\bibitem[{\citenamefont{Abbott et~al.}(2016{\natexlab{b}})\citenamefont{Abbott,
  Abbott, Abbott, Abernathy, Acernese, Ackley, Adams, Adams, Addesso, Adhikari
  et~al.}}]{PhysRevLett.116.241103}
\bibinfo{author}{\bibfnamefont{B.~P.} \bibnamefont{Abbott}},
  \bibinfo{author}{\bibfnamefont{R.}~\bibnamefont{Abbott}},
  \bibinfo{author}{\bibfnamefont{T.~D.} \bibnamefont{Abbott}},
  \bibinfo{author}{\bibfnamefont{M.~R.} \bibnamefont{Abernathy}},
  \bibinfo{author}{\bibfnamefont{F.}~\bibnamefont{Acernese}},
  \bibinfo{author}{\bibfnamefont{K.}~\bibnamefont{Ackley}},
  \bibinfo{author}{\bibfnamefont{C.}~\bibnamefont{Adams}},
  \bibinfo{author}{\bibfnamefont{T.}~\bibnamefont{Adams}},
  \bibinfo{author}{\bibfnamefont{P.}~\bibnamefont{Addesso}},
  \bibinfo{author}{\bibfnamefont{R.~X.} \bibnamefont{Adhikari}},
  \bibnamefont{et~al.} (\bibinfo{collaboration}{LIGO Scientific Collaboration
  and Virgo Collaboration}), \bibinfo{journal}{Phys. Rev. Lett.}
  \textbf{\bibinfo{volume}{116}}, \bibinfo{pages}{241103}
  (\bibinfo{year}{2016}{\natexlab{b}}).

\bibitem[{\citenamefont{Armano et~al.}(2016)\citenamefont{Armano, Audley,
  Auger, Baird, Bassan, Binetruy, Born, Bortoluzzi, Brandt, Caleno
  et~al.}}]{PhysRevLett.116.231101}
\bibinfo{author}{\bibfnamefont{M.}~\bibnamefont{Armano}},
  \bibinfo{author}{\bibfnamefont{H.}~\bibnamefont{Audley}},
  \bibinfo{author}{\bibfnamefont{G.}~\bibnamefont{Auger}},
  \bibinfo{author}{\bibfnamefont{J.~T.} \bibnamefont{Baird}},
  \bibinfo{author}{\bibfnamefont{M.}~\bibnamefont{Bassan}},
  \bibinfo{author}{\bibfnamefont{P.}~\bibnamefont{Binetruy}},
  \bibinfo{author}{\bibfnamefont{M.}~\bibnamefont{Born}},
  \bibinfo{author}{\bibfnamefont{D.}~\bibnamefont{Bortoluzzi}},
  \bibinfo{author}{\bibfnamefont{N.}~\bibnamefont{Brandt}},
  \bibinfo{author}{\bibfnamefont{M.}~\bibnamefont{Caleno}},
  \bibnamefont{et~al.}, \bibinfo{journal}{Phys. Rev. Lett.}
  \textbf{\bibinfo{volume}{116}}, \bibinfo{pages}{231101}
  (\bibinfo{year}{2016}).

\bibitem[{\citenamefont{Widrow and Kaiser}(1993)}]{widrow1993using}
\bibinfo{author}{\bibfnamefont{L.~M.} \bibnamefont{Widrow}} \bibnamefont{and}
  \bibinfo{author}{\bibfnamefont{N.}~\bibnamefont{Kaiser}},
  \bibinfo{journal}{The Astrophys. J.} \textbf{\bibinfo{volume}{416}},
  \bibinfo{pages}{L71} (\bibinfo{year}{1993}).

\bibitem[{\citenamefont{Schaller et~al.}(2014)\citenamefont{Schaller, Becker,
  Ruchayskiy, Boyarsky, and Shaposhnikov}}]{schaller2014new}
\bibinfo{author}{\bibfnamefont{M.}~\bibnamefont{Schaller}},
  \bibinfo{author}{\bibfnamefont{C.}~\bibnamefont{Becker}},
  \bibinfo{author}{\bibfnamefont{O.}~\bibnamefont{Ruchayskiy}},
  \bibinfo{author}{\bibfnamefont{A.}~\bibnamefont{Boyarsky}}, \bibnamefont{and}
  \bibinfo{author}{\bibfnamefont{M.}~\bibnamefont{Shaposhnikov}},
  \bibinfo{journal}{Mon.\ Not.\ Roy.\ Astr.\ Soc.}
  \textbf{\bibinfo{volume}{442}}, \bibinfo{pages}{3073} (\bibinfo{year}{2014}).

\bibitem[{\citenamefont{Widrow}(1997)}]{Widrow1997}
\bibinfo{author}{\bibfnamefont{L.~M.} \bibnamefont{Widrow}},
  \bibinfo{journal}{Phys. Rev. D} \textbf{\bibinfo{volume}{55}},
  \bibinfo{pages}{5997} (\bibinfo{year}{1997}).

\bibitem[{\citenamefont{Widrow and Kaiser}(1997)}]{Davies1997}
\bibinfo{author}{\bibfnamefont{L.~M.} \bibnamefont{Widrow}} \bibnamefont{and}
  \bibinfo{author}{\bibfnamefont{N.}~\bibnamefont{Kaiser}},
  \bibinfo{journal}{The Astrophys. J.} \textbf{\bibinfo{volume}{485}},
  \bibinfo{pages}{484} (\bibinfo{year}{1997}).

\bibitem[{\citenamefont{Poisson and Will}(2014)}]{poisson:gravity}
\bibinfo{author}{\bibfnamefont{E.}~\bibnamefont{Poisson}} \bibnamefont{and}
  \bibinfo{author}{\bibfnamefont{C.}~\bibnamefont{Will}},
  \emph{\bibinfo{title}{Gravity: Newtonian, post-Newtonian, Relativistic}}
  (\bibinfo{publisher}{Cambridge University Press}, \bibinfo{address}{England},
  \bibinfo{year}{2014}).

\bibitem[{\citenamefont{Thorne}(1998)}]{Thorne1998}
\bibinfo{author}{\bibfnamefont{K.~S.} \bibnamefont{Thorne}},
  \bibinfo{journal}{Phys.\ Rev.\ D} \textbf{\bibinfo{volume}{58}},
  \bibinfo{pages}{124031} (\bibinfo{year}{1998}).

\bibitem[{\citenamefont{{Hinderer}}(2008)}]{Hinderer:2008}
\bibinfo{author}{\bibfnamefont{T.}~\bibnamefont{{Hinderer}}},
  \bibinfo{journal}{Astroph. J.} \textbf{\bibinfo{volume}{677}},
  \bibinfo{pages}{1216} (\bibinfo{year}{2008}).

\bibitem[{\citenamefont{Walls and Milburn}(2007)}]{walls2007quantum}
\bibinfo{author}{\bibfnamefont{D.~F.} \bibnamefont{Walls}} \bibnamefont{and}
  \bibinfo{author}{\bibfnamefont{G.~J.} \bibnamefont{Milburn}},
  \emph{\bibinfo{title}{Quantum optics}} (\bibinfo{publisher}{Springer Science
  \& Business Media}, \bibinfo{year}{2007}).

\bibitem[{\citenamefont{Scully and Zubair}(1997)}]{Scully1997}
\bibinfo{author}{\bibfnamefont{M.~O.} \bibnamefont{Scully}} \bibnamefont{and}
  \bibinfo{author}{\bibfnamefont{M.~S.} \bibnamefont{Zubair}},
  \emph{\bibinfo{title}{Quantum Optics}} (\bibinfo{publisher}{Cambridge
  University Press}, \bibinfo{year}{1997}).

\bibitem[{\citenamefont{Kolb and Tkachev}(1994)}]{PhysRevD.50.769}
\bibinfo{author}{\bibfnamefont{E.~W.} \bibnamefont{Kolb}} \bibnamefont{and}
  \bibinfo{author}{\bibfnamefont{I.~I.} \bibnamefont{Tkachev}},
  \bibinfo{journal}{Phys. Rev. D} \textbf{\bibinfo{volume}{50}},
  \bibinfo{pages}{769} (\bibinfo{year}{1994}).

\bibitem[{\citenamefont{Schmid et~al.}(1999)\citenamefont{Schmid, Schwarz, and
  Widerin}}]{PhysRevD.59.043517}
\bibinfo{author}{\bibfnamefont{C.}~\bibnamefont{Schmid}},
  \bibinfo{author}{\bibfnamefont{D.~J.} \bibnamefont{Schwarz}},
  \bibnamefont{and} \bibinfo{author}{\bibfnamefont{P.}~\bibnamefont{Widerin}},
  \bibinfo{journal}{Phys. Rev. D} \textbf{\bibinfo{volume}{59}},
  \bibinfo{pages}{043517} (\bibinfo{year}{1999}).

\bibitem[{\citenamefont{Silk and Stebbins}(1993)}]{silk1992dmc}
\bibinfo{author}{\bibfnamefont{J.}~\bibnamefont{Silk}} \bibnamefont{and}
  \bibinfo{author}{\bibfnamefont{A.}~\bibnamefont{Stebbins}},
  \bibinfo{journal}{Astrophys. J.} \textbf{\bibinfo{volume}{411}},
  \bibinfo{pages}{439} (\bibinfo{year}{1993}).

\bibitem[{\citenamefont{Bertschinger}(1985)}]{bertschinger1985self}
\bibinfo{author}{\bibfnamefont{E.}~\bibnamefont{Bertschinger}},
  \bibinfo{journal}{The Astrophysical Journal Supplement Series}
  \textbf{\bibinfo{volume}{58}}, \bibinfo{pages}{39} (\bibinfo{year}{1985}).

\bibitem[{\citenamefont{Aslanyan et~al.}(2016)\citenamefont{Aslanyan, Price,
  Adams, Bringmann, Clark, Easther, Lewis, and Scott}}]{PhysRevLett.117.141102}
\bibinfo{author}{\bibfnamefont{G.}~\bibnamefont{Aslanyan}},
  \bibinfo{author}{\bibfnamefont{L.~C.} \bibnamefont{Price}},
  \bibinfo{author}{\bibfnamefont{J.}~\bibnamefont{Adams}},
  \bibinfo{author}{\bibfnamefont{T.}~\bibnamefont{Bringmann}},
  \bibinfo{author}{\bibfnamefont{H.~A.} \bibnamefont{Clark}},
  \bibinfo{author}{\bibfnamefont{R.}~\bibnamefont{Easther}},
  \bibinfo{author}{\bibfnamefont{G.~F.} \bibnamefont{Lewis}}, \bibnamefont{and}
  \bibinfo{author}{\bibfnamefont{P.}~\bibnamefont{Scott}},
  \bibinfo{journal}{Phys. Rev. Lett.} \textbf{\bibinfo{volume}{117}},
  \bibinfo{pages}{141102} (\bibinfo{year}{2016}).

\bibitem[{\citenamefont{Berezinsky et~al.}(2006)\citenamefont{Berezinsky,
  Dokuchaev, and Eroshenko}}]{PhysRevD.73.063504}
\bibinfo{author}{\bibfnamefont{V.}~\bibnamefont{Berezinsky}},
  \bibinfo{author}{\bibfnamefont{V.}~\bibnamefont{Dokuchaev}},
  \bibnamefont{and}
  \bibinfo{author}{\bibfnamefont{Y.}~\bibnamefont{Eroshenko}},
  \bibinfo{journal}{Phys. Rev. D} \textbf{\bibinfo{volume}{73}},
  \bibinfo{pages}{063504} (\bibinfo{year}{2006}).

\bibitem[{\citenamefont{Berezinsky et~al.}(2013)\citenamefont{Berezinsky,
  Dokuchaev, and Eroshenko}}]{Berezinsky2013}
\bibinfo{author}{\bibfnamefont{V.}~\bibnamefont{Berezinsky}},
  \bibinfo{author}{\bibfnamefont{V.}~\bibnamefont{Dokuchaev}},
  \bibnamefont{and}
  \bibinfo{author}{\bibfnamefont{Y.}~\bibnamefont{Eroshenko}},
  \bibinfo{journal}{Journal of Cosmology and Astroparticle Physics}
  \textbf{\bibinfo{volume}{2013}}, \bibinfo{pages}{059} (\bibinfo{year}{2013}).

\bibitem[{\citenamefont{Ali-Ha\"{\i}moud
  et~al.}(2016)\citenamefont{Ali-Ha\"{\i}moud, Kovetz, and
  Silk}}]{PhysRevD.93.043508}
\bibinfo{author}{\bibfnamefont{Y.}~\bibnamefont{Ali-Ha\"{\i}moud}},
  \bibinfo{author}{\bibfnamefont{E.~D.} \bibnamefont{Kovetz}},
  \bibnamefont{and} \bibinfo{author}{\bibfnamefont{J.}~\bibnamefont{Silk}},
  \bibinfo{journal}{Phys. Rev. D} \textbf{\bibinfo{volume}{93}},
  \bibinfo{pages}{043508} (\bibinfo{year}{2016}).

\bibitem[{\citenamefont{Abbott et~al.}(2009)\citenamefont{Abbott, Abbott,
  Adhikari, Ajith, Allen, Allen, Amin, Anderson, Anderson, Arain
  et~al.}}]{LIGOreport}
\bibinfo{author}{\bibfnamefont{B.~P.} \bibnamefont{Abbott}},
  \bibinfo{author}{\bibfnamefont{R.}~\bibnamefont{Abbott}},
  \bibinfo{author}{\bibfnamefont{R.}~\bibnamefont{Adhikari}},
  \bibinfo{author}{\bibfnamefont{P.}~\bibnamefont{Ajith}},
  \bibinfo{author}{\bibfnamefont{B.}~\bibnamefont{Allen}},
  \bibinfo{author}{\bibfnamefont{G.}~\bibnamefont{Allen}},
  \bibinfo{author}{\bibfnamefont{R.~S.} \bibnamefont{Amin}},
  \bibinfo{author}{\bibfnamefont{S.~B.} \bibnamefont{Anderson}},
  \bibinfo{author}{\bibfnamefont{W.~G.} \bibnamefont{Anderson}},
  \bibinfo{author}{\bibfnamefont{M.~A.} \bibnamefont{Arain}},
  \bibnamefont{et~al.} (\bibinfo{collaboration}{LIGO Scientific
  Collaboration}), \bibinfo{journal}{Rep. Prog. Phys.}
  \textbf{\bibinfo{volume}{72}}, \bibinfo{pages}{076901}
  (\bibinfo{year}{2009}).

\bibitem[{\citenamefont{Amaro-Seoane et~al.}(2013)\citenamefont{Amaro-Seoane,
  Aoudia, Babak, Bin{\'e}truy, Berti, Boh{\'e}, Caprini, Colpi, Cornish,
  Danzmann et~al.}}]{amaro2012elisa}
\bibinfo{author}{\bibfnamefont{P.}~\bibnamefont{Amaro-Seoane}},
  \bibinfo{author}{\bibfnamefont{S.}~\bibnamefont{Aoudia}},
  \bibinfo{author}{\bibfnamefont{S.}~\bibnamefont{Babak}},
  \bibinfo{author}{\bibfnamefont{P.}~\bibnamefont{Bin{\'e}truy}},
  \bibinfo{author}{\bibfnamefont{E.}~\bibnamefont{Berti}},
  \bibinfo{author}{\bibfnamefont{A.}~\bibnamefont{Boh{\'e}}},
  \bibinfo{author}{\bibfnamefont{C.}~\bibnamefont{Caprini}},
  \bibinfo{author}{\bibfnamefont{M.}~\bibnamefont{Colpi}},
  \bibinfo{author}{\bibfnamefont{N.~J.} \bibnamefont{Cornish}},
  \bibinfo{author}{\bibfnamefont{K.}~\bibnamefont{Danzmann}},
  \bibnamefont{et~al.}, \bibinfo{journal}{GW Notes}
  \textbf{\bibinfo{volume}{6}}, \bibinfo{pages}{4} (\bibinfo{year}{2013}).

\bibitem[{\citenamefont{Macedo et~al.}(2013{\natexlab{a}})\citenamefont{Macedo,
  Pani, Cardoso, and Crispino}}]{macedo2013into}
\bibinfo{author}{\bibfnamefont{C.~F.} \bibnamefont{Macedo}},
  \bibinfo{author}{\bibfnamefont{P.}~\bibnamefont{Pani}},
  \bibinfo{author}{\bibfnamefont{V.}~\bibnamefont{Cardoso}}, \bibnamefont{and}
  \bibinfo{author}{\bibfnamefont{L.~C.} \bibnamefont{Crispino}},
  \bibinfo{journal}{The Astrophysical Journal} \textbf{\bibinfo{volume}{774}},
  \bibinfo{pages}{48} (\bibinfo{year}{2013}{\natexlab{a}}).

\bibitem[{\citenamefont{Yoshida et~al.}(1994)\citenamefont{Yoshida, Eriguchi,
  and Futamase}}]{Yoshida1994}
\bibinfo{author}{\bibfnamefont{S.}~\bibnamefont{Yoshida}},
  \bibinfo{author}{\bibfnamefont{Y.}~\bibnamefont{Eriguchi}}, \bibnamefont{and}
  \bibinfo{author}{\bibfnamefont{T.}~\bibnamefont{Futamase}},
  \bibinfo{journal}{Phys. Rev. D} \textbf{\bibinfo{volume}{50}},
  \bibinfo{pages}{6235} (\bibinfo{year}{1994}).

\bibitem[{\citenamefont{Macedo et~al.}(2013{\natexlab{b}})\citenamefont{Macedo,
  Pani, Cardoso, and Crispino}}]{PhysRevD.88.064046}
\bibinfo{author}{\bibfnamefont{C.~F.~B.} \bibnamefont{Macedo}},
  \bibinfo{author}{\bibfnamefont{P.}~\bibnamefont{Pani}},
  \bibinfo{author}{\bibfnamefont{V.}~\bibnamefont{Cardoso}}, \bibnamefont{and}
  \bibinfo{author}{\bibfnamefont{L.~C.~B.} \bibnamefont{Crispino}},
  \bibinfo{journal}{Phys. Rev. D} \textbf{\bibinfo{volume}{88}},
  \bibinfo{pages}{064046} (\bibinfo{year}{2013}{\natexlab{b}}).

\bibitem[{\citenamefont{Stone et~al.}(2013)\citenamefont{Stone, Sari, and
  Loeb}}]{stone2013consequences}
\bibinfo{author}{\bibfnamefont{N.}~\bibnamefont{Stone}},
  \bibinfo{author}{\bibfnamefont{R.}~\bibnamefont{Sari}}, \bibnamefont{and}
  \bibinfo{author}{\bibfnamefont{A.}~\bibnamefont{Loeb}},
  \bibinfo{journal}{Mon.\ Not.\ Roy.\ Astr.\ Soc.}
  \textbf{\bibinfo{volume}{435}}, \bibinfo{pages}{1809} (\bibinfo{year}{2013}).

\end{thebibliography}

\end{document}